\documentclass[iop]{emulateapj}
\usepackage{graphicx}
\usepackage{amssymb}
\usepackage{natbib}

\newcommand{\VOR}{\rm vor}
\newcommand{\VORS}{\overline{\rm vor \sin \theta}}
\newcommand{\DIV}{\rm div}

\newcommand{\UFIELD}{\overline{\vert B_r \vert}}
\newcommand{\DTMEAN}{\overline{\delta \tau_{\rm mn}}}
\newcommand{\DTOI}{\overline{\delta \tau_{\rm oi}}}

\newcommand{\DTOUT}{\overline{\delta \tau_{\rm out}}}
\newcommand{\DTIN}{\overline{\delta \tau_{\rm in}}}
\newcommand{\DTINC}{\overline{\delta \tau_{\rm in} \cos \theta}}

\newcommand{\DTXC}{\overline{\delta \tau_x \cos \theta}}

\newcommand{\DTYS}{\overline{\delta \tau_y \sin \theta}}
\defcitealias{Lekaetal2013}{Paper I}
\defcitealias{Birchetal2013}{Paper II}

\begin{document}

\title{Helioseismology of Pre-Emerging Active Regions III: Statistical Analysis}
\author{G.~Barnes, A.C.~Birch\altaffilmark{1}, K.D.~Leka and D.C.~Braun}
\affil{NorthWest Research Associates, Boulder, CO 80301 USA}
\email{graham@nwra.com}

\altaffiltext{1}{Max-Planck-Institut f\"{u}r Sonnensystemforschung, D-37191
Katlenburg-Lindau, Germany}


\shorttitle{Emerging Active Regions III}
\shortauthors{Barnes et al.}

\begin{abstract}

The subsurface properties of active regions prior to their appearance at the
solar surface may shed light on the process of active region formation.
Helioseismic holography has been applied to samples taken from two populations
of regions on the Sun (pre-emergence and without emergence), each sample having
over 100 members, that were selected to minimize systematic bias, as described
in Paper I (Leka et al., 2012).  Paper II (Birch et al., 2012) showed that
there are statistically significant signatures in the average helioseismic
properties that precede the formation of an active region.  This paper
describes a more detailed analysis of the samples of pre-emergence regions and
regions without emergence, based on discriminant analysis.  The property that
is best able to distinguish the populations is found to be the surface magnetic
field, even a day before the emergence time.  However, after accounting for the
correlations between the surface field and the quantities derived from
helioseismology, there is still evidence of a helioseismic precursor to active
region emergence that is present for at least a day prior to emergence.

\end{abstract}

\maketitle

\section{Introduction}\label{sec:introduction}
 
Models for the formation of solar active regions (AR) tend to fall into one of
several classes, largely dependent on the volume in which strong magnetic
fields are generated.  In one of these, magnetic flux tubes generated near the
base of the convection zone become buoyant and rise through the convection zone
\citep[for a review see][]{Fan2009}, with an active region emerging when a flux
tube passes through the solar surface.  Another possibility is that ARs are
formed as the result of the coalescence of magnetic fields generated in the
bulk of the convection zone or near the solar surface \citep[][and references
therein]{Brandenburg2005}.  Each of these scenarios has a distinct signature in
the velocity of the plasma in the convection zone.  Local helioseismology
\citep{GizonBirch2005,Gizonetal2010} potentially can be used to determine the
subsurface dynamics associated with AR formation, and thus could provide
evidence for or against either of these theories.  This would also indirectly
shed light on the location of the solar dynamo. 

Previous studies of AR formation using local helioseismology have tended to
focus on only a small number of regions
\citep[e.g.,][]{Braun1995,Changetal1999,Jensenetal2001,ZharkovThompson2008,Kosovichev2009,Hartlepetal2011,Ilonidisetal2011}.
The small number of regions considered and the lack of a control group of areas
of Sun where no active region was emerging make it difficult to identify any
subsurface properties {\em unique} to the emergence of active regions.  The
exception to this is the study of \cite{Kommetal2009,Kommetal2011} that
considered subsurface flows of a large sample of existing active regions
undergoing episodes of magnetic flux emergence, compared with a control group
of active regions that had comparatively constant flux.  However, this study
did not include the pre-emergence stage of active region formation.

The present study is based on applying helioseismic holography
\citep{LindseyBraun2000} to samples of over 100 areas of Sun where an active
region subsequently emerged, and an equal number where no active region
emerged. The selection of these regions was described in \citet{Lekaetal2013}
(Paper I), while an initial analysis of the travel-times inferred from
helioseismic holography, focusing on the average travel-time shifts, was
presented in \citet{Birchetal2013} (Paper II).  It was found that there are
statistically significant differences in the average travel-times, as well as
in the surface magnetic flux, between the samples of pre-emergence areas and
quiet sun areas.  These included a reduction in the mean travel-time shift of a
few tenths of a second, as well as spatially antisymmetric features in both the
east-west and north-south travel-time differences.  The antisymmetric features
are qualitatively consistent with what would be expected from a flow converging
on the site of emergence, although it appears that it is not a simple
converging flow.  One possible interpretation is that emergence preferentially
occurs at the boundaries between supergranules.  In this scenario, the
emergence is not at the center of a converging flow, but between neighboring
diverging flows.  This could also account for the difference in the surface
magnetic field, as flux tends to concentrate in the boundaries between
supergranules. 

As interesting as what was found in \citetalias{Birchetal2013} is what was not
found: any signature of a strong retrograde flow, or any travel-time shifts
greater than of order one second.  Simulations of rising flux tubes
\citep[e.g.,][]{Fan2008} predict retrograde flows with magnitudes of order
100\,m s${}^{-1}$, while \cite{Ilonidisetal2011} found mean travel-time
reductions of order 10\,s.  In both these cases, the results are much larger
effects than were found in \citetalias{Birchetal2013}, but for layers in the
sun significantly below the roughly 20\,Mm maximum depth considered here.  Thus
it may be that there are significant changes in the emergence process between
depths of approximately 60\,Mm and 20\,Mm.  

In the present paper, we briefly review the selection of the data, and the
analysis performed in the previous papers in this series before proceeding to
an analysis of the data based on discriminant analysis.  We use the full
distribution of the travel-time shifts to determine the relative ability of
different parameters to discriminate between the samples of pre-emergence and
non-emergence.  We compare the ability of a measure of the surface magnetic
field to distinguish the samples with the ability of the helioseismic
parameters, and examine how the surface field is influencing the helioseismic
parameters.  An important caveat is that here, as in
\citetalias{Birchetal2013}, care should be taken in interpreting the nature of
the holography travel-time shifts. For example, without modeling, the variation
of depth of any flows or other perturbations producing the shifts is not known.

\section{The Data and Helioseismic Analysis}\label{sec:data}

The overall design of this study, including the data selection, preparation and
treatment, was presented in Paper I.  In brief, samples from two populations are
considered, ``Pre-Emergence'' targets (PE) that track a $32\degr \times
32\degr$ patch of the Sun prior to the emergence of a NOAA-numbered active
region, and ``Non-Emergence'' targets (NE) selected for lack of emergence and
lack of strong fields in the central portions of the tracked patch.  The PE
sample size comprises 107 targets obtained between 2001--2007, matched to 107
NE targets drawn from an initially larger sample, and selected further to match
the PE distributions in time and observing location on the disk.  The emergence
time was determined using MDI 96-minute cadence observations of the line of
sight magnetic field, no selection was made for minimum-size of the numbered
NOAA regions that result and limits were placed to avoid extreme observing
angles.  

Data for the helioseismology originate from the Global Oscillations Network
Group project \citep[GONG;][]{gong,gong1}; a full GONG day (1664 min) of data
prior to the emergence time was tracked, and divided into five time intervals,
each 6.4\,hr long but starting every 5.3\,hr, with just over an hour overlap
between them.  Table~\ref{tbl:samplesize} shows how many NE/PE had acceptable
duty cycle ($>80$\%) for each time interval; once duty cycle is accounted for,
the sample sizes are not equal.  

\begin{deluxetable}{cccc}
\tablecolumns{4}
\tablecaption{Sample Sizes \label{tbl:samplesize}}
\tablewidth{0pt}
\tablehead{
\colhead{Time Interval} & \colhead{time before emergence} & \colhead{\#NE} 
& \colhead{\#PE} \cr
& (hr) & & }
\startdata
TI-0 &$t_0-24.5$ & 81 & 89 \cr
TI-1 &$t_0-19.2$ & 85 & 88 \cr
TI-2 &$t_0-13.9$ & 85 & 89 \cr
TI-3 & $t_0-8.5$ & 82 & 87 \cr
TI-4 & $t_0-3.2$ & 83 & 86 \cr
\enddata
\end{deluxetable}

We measured wave travel-times from each of the time intervals of the GONG data
using surface-focusing helioseismic holography
\citep{LindseyBraun2000,GizonBirch2005}, a technique very similar to
time-distance helioseismology \citep{Duvalletal1993}.  In particular, the GONG
Dopplergrams were first tracked and Postel projected, then phase speed filters
were applied.  The filters, described in Table~1 of \citet{Couvidatetal2005},
isolate waves with particular ranges in lower turning points; these filters
cover the range in lower turning point depths from about 1.4\,Mm (filter TD1)
to about 23.3\,Mm (filter TD11).  The full list of depths is given in Table~1
of \citetalias{Birchetal2013}.  After filtering, center-annulus and
center-quadrant local-control correlations were used to measure travel-time
shifts.  From these, travel-time differences and proxies for the vertical
component of the flow vorticity and the horizontal flow divergence were
constructed.  The result was, for each time interval and each region, a spatial
map of the travel-times listed in Table~\ref{tbl:vars}. 

\begin{deluxetable}{ccccc}
\tablecolumns{5}
\tablewidth{0pc}
\tablecaption{Helioseismology and magnetic variables \label{tbl:vars}}
\tablehead{
\colhead{variable} & \colhead{description}}
\startdata
$\delta \tau_x$ & east-west travel-time difference \\
$\delta \tau_y$ & north-south travel-time difference \\
$\delta \tau_{\rm in}$ & annulus-to-center travel-time shift \\
$\delta \tau_{\rm out}$ & center-to-annulus travel-time shift \\
$\delta \tau_{\rm oi}$ & ``out minus in'' travel-time difference
$\delta \tau_{\rm out} - \delta \tau_{\rm in}$ \\
$\delta \tau_{\rm mn}$ & mean travel-time shift 
$[\delta \tau_{\rm out} + \delta \tau_{\rm in}]/2$ \\
$\VOR$ & vertical component of vorticity: $\partial_x \delta \tau_y - \partial_y \delta \tau_x$ \\
$\DIV$ & horizontal flow divergence: $\partial_x \delta \tau_x + \partial_y \delta \tau_y$ \\
$B_r$ & radial component of (potential) magnetic field \\
\enddata
\tablecomments{All measures of the magnetic field are averaged over the
corresponding time interval.}
\end{deluxetable}

In order to reduce the spatial maps to a small number of parameters
characterizing each region during each time interval, each travel-time map was
spatially averaged over a 45.5\,Mm disk, centered at the emergence location for
PE.  As in \citetalias{Birchetal2013}, three weightings were used in the
averaging: a uniform weighting, and $\sin \theta$ and $\cos \theta$ weightings,
where $\theta$ is the angle measured counter-clockwise from the direction of
solar rotation (the $+\hat{\bf x}$ direction).  Using this combination of
weighting factors makes the analysis sensitive to both spatially symmetric and
antisymmetric features in the travel-time maps.  To be consistent with
\citetalias{Birchetal2013}, we will continue to denote the spatial average with
an overline, but note that only spatial averages are considered here. 

The accompanying magnetic data derive from MDI observations: a potential field
was calculated that matches the observed line-of-sight component provided by
MDI, hence providing the potential-field approximation of the radial field
present over the course of the GONG data, for comparison with the results of
the helioseismology.  The absolute value of the radial magnetic field was
spatially averaged over the same 45.5\,Mm disk as the travel-times, and
temporally averaged over each time interval.  Note that after accounting for
duty cycle, the magnetic field variables have different sample sizes than the
seismology variables because they were computed from MDI data; only time
interval 0 has a sample size not equal to 107, where the NE sample is reduced
to 106.

\section{Statistical Tests: Discriminant Analysis and Skill Scores}\label{sec:stats}

The analysis presented in Paper II suggests that helioseismic holography is
able to detect a signature prior to the emergence time as defined in
\citetalias{Lekaetal2013}.  To quantify this ability, and in particular to
determine whether there is any more information available from the holographic
signatures than there is from direct measurements of the surface magnetic
field, discriminant analysis \citep[e.g.,][]{ken83} was used.  This technique
classifies a measurement as belonging to the group with the highest probability
density.  Provided the probability density is estimated accurately, it
maximizes the overall rate of correct classification.  In this case, the two
groups are the PE and NE regions, and a region would be classified as emerging
whenever the probability density estimate for the emerging regions exceeds the
probability density estimate for quiet regions, for the specified property of
the new region.  

For the results presented here, the probability density was estimated using a
kernel method with the Epanechnikov kernel and the smoothing parameter set
based on its optimum value for a normal distribution \citep{Silverman86}.  For
the average unsigned flux, which is a positive definite quantity, the
probability density of the logarithm was estimated.  This ensures that the
density estimate is zero for values of the flux less than zero, and better
captures the typical tail to high values.  Example density estimates are 
presented in \S\ref{sec:results}. 

At any randomly selected point on the solar disk, the probability of an active
region emerging in a one day window is extremely small.  However, for this
analysis, the prior probabilities for pre-emergence and non-emergence were set
equal.  Thus, we are not truly testing the ability of the parameters to {\it
predict} the emergence of an active region, but rather we are testing whether
there is a signal of emergence in the seismic analysis. The advantage to this
choice is that it avoids the problem that the presence of even quite a strong
signal can be masked by an extremely small prior probability. 

The first step in quantifying the performance of the discriminant analysis was
to construct a contingency table, as shown in Table~\ref{tbl:contingency}.
From a contingency table, there are many ways to quantify the performance of a
classification scheme.  Because prior probabilities are assumed to be equal,
but the sample sizes are not equal after accounting for the duty cycle (see
\S\ref{sec:data} and Table~\ref{tbl:samplesize}), we use the Peirce skill score
\citep{Peirce1884}, also known as the true skill score, or Hanssen and
Kuipers's discriminant \cite[see][for a comparison of this with other skill
scores]{Woodcock1976}. It is given by
\begin{eqnarray}
{\rm PSS} &=& {n_{pp} \over n_p} - {n_{np} \over n_n},
\end{eqnarray}
where $n_{pp}$ is the number of regions that were classified by the
discriminant analysis to be emergences and did emerge, $n_p$ is the number of
PE regions, $n_{np}$ is the number of regions that were classified by the
discriminant analysis to be non-emergences but did emerge, and $n_n$ is the
number of NE regions.  As expressed above, the Peirce skill score is the
probability of detection (hit rate) minus the probability of false detection
(false alarm rate).  Changing the sample size of events or non-events does not
change this score provided the rates have been accurately estimated.  Positive
values of this skill score indicate improvement of the forecasts over both
uniform (unskilled) forecasts and random forecasts \citep{Woodcock1976}, with a
maximum score of $1.0$ for perfect forecasting, while negative scores indicate
worse performance than uniform or random forecasts. 

\begin{deluxetable}{crrrr}
\tablecolumns{4}
\tablewidth{0pc}
\tablecaption{Contingency Table 
\label{tbl:contingency}}
\tablehead{
\colhead{} & \colhead{} & \multicolumn{2}{c}{Classified}}
\startdata
         &    & PE       & NE \\
Observed & PE & $n_{pp}$ & $n_{np}$ \\
         & NE & $n_{pn}$ & $n_{nn}$ \\
\enddata
\end{deluxetable}

To further confirm the independence of the results on the varying sample sizes,
the analysis was repeated using only the subset of regions that have good duty
cycles for all time intervals.  Sample results of this investigation are shown
in appendix~\ref{sec:duty}.  The main result of using this subset is to
increase the uncertainty estimates, which is a consequence of the sample sizes
being substantially reduced to 45/48 for NE/PE; the significance of the results
is thus reduced, but the interpretations remain the same.  

An unbiased estimate of the skill score with an error estimate was obtained by
using cross-validation \citep{Hills66} and a bootstrap approach
\citep[e.g.,][]{EfronGong1983}.  For each sample (PE and NE), a bootstrap
sample was constructed by drawing with replacement from the full sample.  That
is, a PE (NE) region was selected at random from the full set of PE (NE)
regions, and this was repeated $n_p$ ($n_n$) times to construct a bootstrap
sample.  Because all the draws are from the full sample, the same region may be
drawn more than once, or not at all.

To remove bias, each member of the bootstrap sample was classified by using the
remaining $n-1$ points to determine the probability density at the removed
point, and repeating for all $n$ points in the sample, from which a contingency
table and skill score were constructed.  This was repeated for 1000 bootstrap
samples, with the mean and standard deviation of the resulting skill scores
used to estimate the skill score value and error.  

\section{Results}\label{sec:results}

To determine the variables with the greatest ability to distinguish between PE
and NE regions, the unbiased estimate of the Peirce skill score and its
uncertainty from nonparametric discriminant analysis were computed for all the
variables in Table~\ref{tbl:vars} for each time interval and phase speed filter
(except the magnetic field, for which no filters were used).
Table~\ref{tbl:peirce_boot_npda_equal_n1_seg} lists all the variables with a
skill score of more than 0.27 for each time interval.\footnote{Skill scores for
all the variables are available as a Machine Readable Table in the online
edition of this paper.} The results of a Monte-Carlo experiment
(appendix~\ref{sec:MC}) show that it is very likely that a variable with a
skill score of greater than 0.27 can truly differentiate between the
populations.  This cutoff is arbitrary in the sense that there are variables
just below the cutoff that have essentially the same ability to discriminate NE
from PE regions as variables that appear in the table.  However, skill scores
up to at least 0.2 can reasonably be expected due to chance for variables that
have no difference between the populations.  Thus the threshold chosen means
that the variables in the table are ones that are very likely to have a real
ability to discriminate; others with real ability may be excluded. 

\begin{deluxetable}{crc}
\tablecolumns{3}
\tablewidth{0pc}
\tablecaption{Best Performing Variables\label{tbl:peirce_boot_npda_equal_n1_seg}}
\tablehead{
\colhead{variable} & \colhead{depth} & \colhead{Peirce SS} \\
\colhead{} & \colhead{(Mm)} & \colhead{}}
\startdata
\cutinhead{TI-0: time=$t_0-24.5$\,hr}
$\overline{\vert B_r \vert}$ & -- & $ 0.38\pm 0.07$ \\
$\overline{\delta \tau_y \sin \theta}^\dagger$ &  2.2 & $ 0.34\pm 0.08$ \\
$\overline{\delta \tau_y \sin \theta}^\dagger$ &  3.2 & $ 0.32\pm 0.08$ \\
$\overline{\VOR \sin \theta}^\dagger$ & 15.7 & $ 0.29\pm 0.09$ \\
$\overline{\VOR \sin \theta}^\dagger$ &  6.2 & $ 0.27\pm 0.09$ \\
\cutinhead{TI-1: time=$t_0-19.2$\,hr}
$\overline{\vert B_r \vert}$ & -- & $ 0.38\pm 0.07$ \\
$\overline{\delta \tau_y \sin \theta}$ &  2.2 & $ 0.29\pm 0.08$ \\
$\overline{\delta \tau_{\rm oi}}$ & 11.4 & $ 0.28\pm 0.08$ \\
$\overline{\delta \tau_{\rm in} \cos \theta}$ &  9.5 & $ 0.28\pm 0.09$ \\
\cutinhead{TI-2: time=$t_0-13.9$\,hr}
$\overline{\vert B_r \vert}$ & -- & $ 0.41\pm 0.07$ \\
$\overline{\delta \tau_x \cos \theta}$ & 11.4 & $ 0.30\pm 0.09$ \\
$\overline{\VOR \sin \theta}^\dagger$ &  6.2 & $ 0.29\pm 0.08$ \\
$\overline{\delta \tau_x \cos \theta}^\dagger$ &  3.2 & $ 0.29\pm 0.08$ \\
$\overline{\delta \tau_{\rm mn}}$ &  6.2 & $ 0.27\pm 0.07$ \\
\cutinhead{TI-3: time= $t_0-8.5$\,hr}
$\overline{\vert B_r \vert}$ & -- & $ 0.41\pm 0.07$ \\
\cutinhead{TI-4: time= $t_0-3.2$\,hr}
$\overline{\vert B_r \vert}$ & -- & $ 0.49\pm 0.06$ \\
$\overline{\delta \tau_{\rm mn}}^\dagger$ &  9.5 & $ 0.44\pm 0.08$ \\
$\overline{\delta \tau_{\rm in}}^\dagger$ &  9.5 & $ 0.36\pm 0.08$ \\
$\overline{\delta \tau_{\rm out}}^\dagger$ &  9.5 & $ 0.36\pm 0.07$ \\
$\overline{\delta \tau_{\rm mn}}^\dagger$ & 20.9 & $ 0.36\pm 0.07$ \\
$\overline{\delta \tau_{\rm mn}}^\dagger$ & 23.3 & $ 0.34\pm 0.08$ \\
$\overline{\delta \tau_{\rm mn}}$ &  6.2 & $ 0.34\pm 0.07$ \\
$\overline{\delta \tau_{\rm in}}$ & 20.9 & $ 0.32\pm 0.08$ \\
$\overline{\delta \tau_{\rm out}}$ & 15.7 & $ 0.30\pm 0.08$ \\
$\overline{\delta \tau_{\rm in}}$ & 23.3 & $ 0.30\pm 0.08$ \\
$\overline{\delta \tau_{\rm mn}}$ & 11.4 & $ 0.30\pm 0.08$ \\
$\overline{\delta \tau_{\rm mn}}$ & 13.3 & $ 0.29\pm 0.08$ \\
$\overline{\delta \tau_{\rm in}}$ & 11.4 & $ 0.28\pm 0.08$ \\
$\overline{\delta \tau_x \cos \theta}^\dagger$ &  3.2 & $ 0.28\pm 0.08$ \\
$\overline{\delta \tau_{\rm mn}}$ & 15.7 & $ 0.28\pm 0.09$ \\
$\overline{\delta \tau_{\rm out}}$ &  2.2 & $ 0.28\pm 0.07$ \\
$\overline{\delta \tau_{\rm in}}$ & 13.3 & $ 0.27\pm 0.08$ \\
$\overline{\delta \tau_{\rm in}}$ & 15.7 & $ 0.27\pm 0.08$ \\
$\overline{\delta \tau_y \sin \theta}^\dagger$ &  2.2 & $ 0.27\pm 0.10$ \\
$\overline{\VOR \sin \theta}^\dagger$ &  6.2 & $ 0.27\pm 0.08$ \\
\enddata
\tablecomments{Depth refers to the lower turning point of the waves in the
filter used.  Time is relative to the emergence time $t_0$.  Variables marked with a ${}^\dagger$
also appear in Table~\ref{tbl:subsetpeirce_boot_npda_equal_n1_seg}, and have
significant ability to discriminate the populations after controlling for
$\UFIELD$, as discussed in \S\ref{sec:fluxmatch}. A version of
Table~\ref{tbl:peirce_boot_npda_equal_n1_seg} containing all the variables
considered is published in the electronic edition of Barnes et al.~(2013).}
\end{deluxetable}

In each time interval, the variable that is best able to distinguish the PE
from the NE regions is the average unsigned field, $\UFIELD$.  In the time
interval immediately prior to emergence (centered 3.2\,hr before the emergence
time), the mean travel-time shift in a variety of filters shows significant
ability to distinguish PE from NE, as do the center-to-annulus and
annulus-to-center travel-times.  In almost all the time intervals,
antisymmetric-weighted averages of the east-west ($\DTXC$) and the north-south
($\DTYS$) travel-time differences measured in filters with shallow lower
turning points appear.  These same measures are highlighted in Figures~4 and 5
of \citetalias{Birchetal2013}, and are interpreted as being consistent with a
converging flow.  The other variable that appears in multiple time intervals is
$\VORS$ in filters with a moderate depth lower turning point.  Several other
variable and filter combinations appear in only one time interval, such as the
difference between the center-to-annulus and the annulus-to-center travel-time,
$\DTOI$, at moderate depth, and an antisymmetric average of the
annulus-to-center travel-time, $\DTINC$, at moderate depth. 

\subsection{The Average Unsigned Magnetic Field}

\begin{figure*}
\plottwo{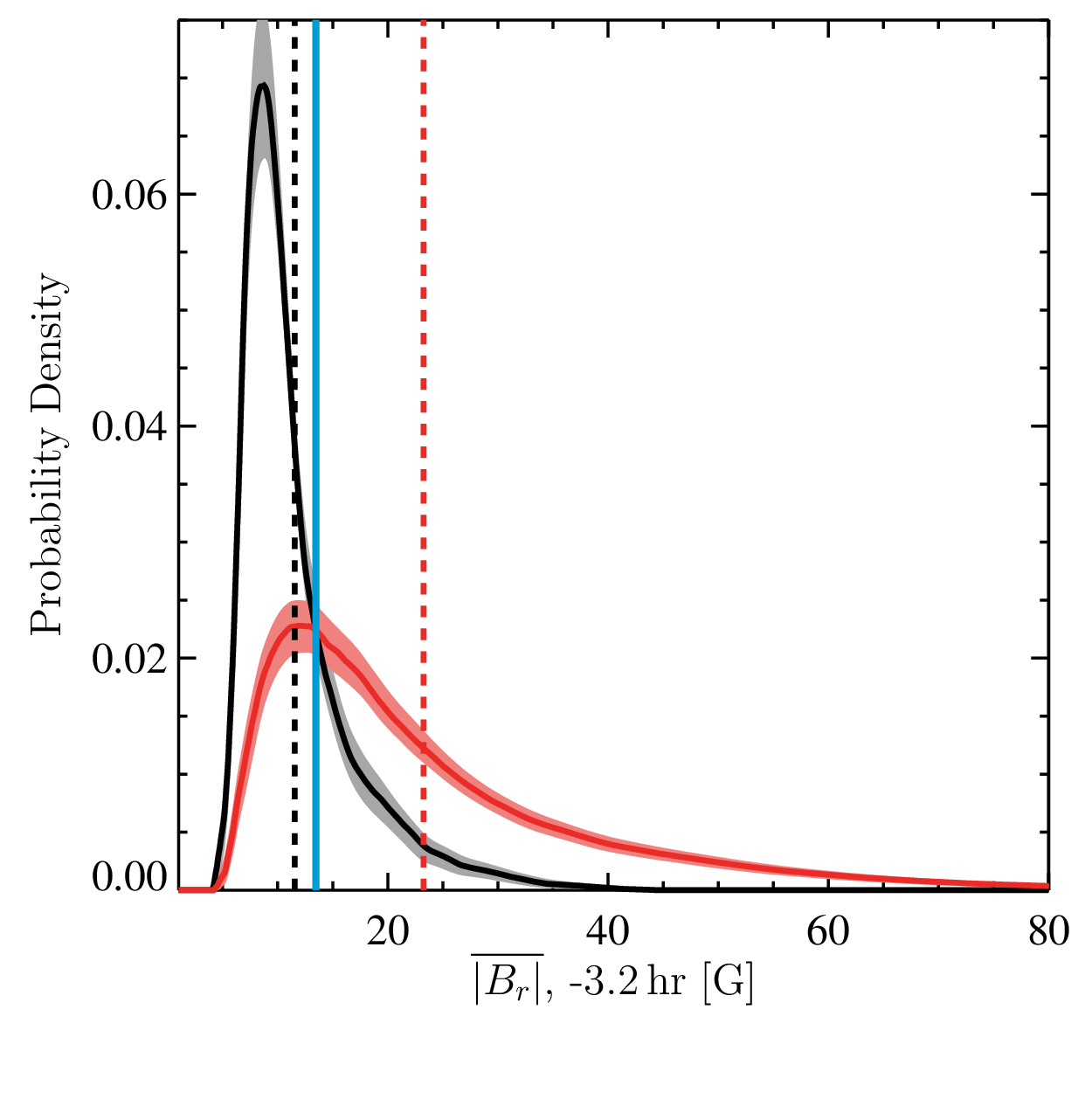}
{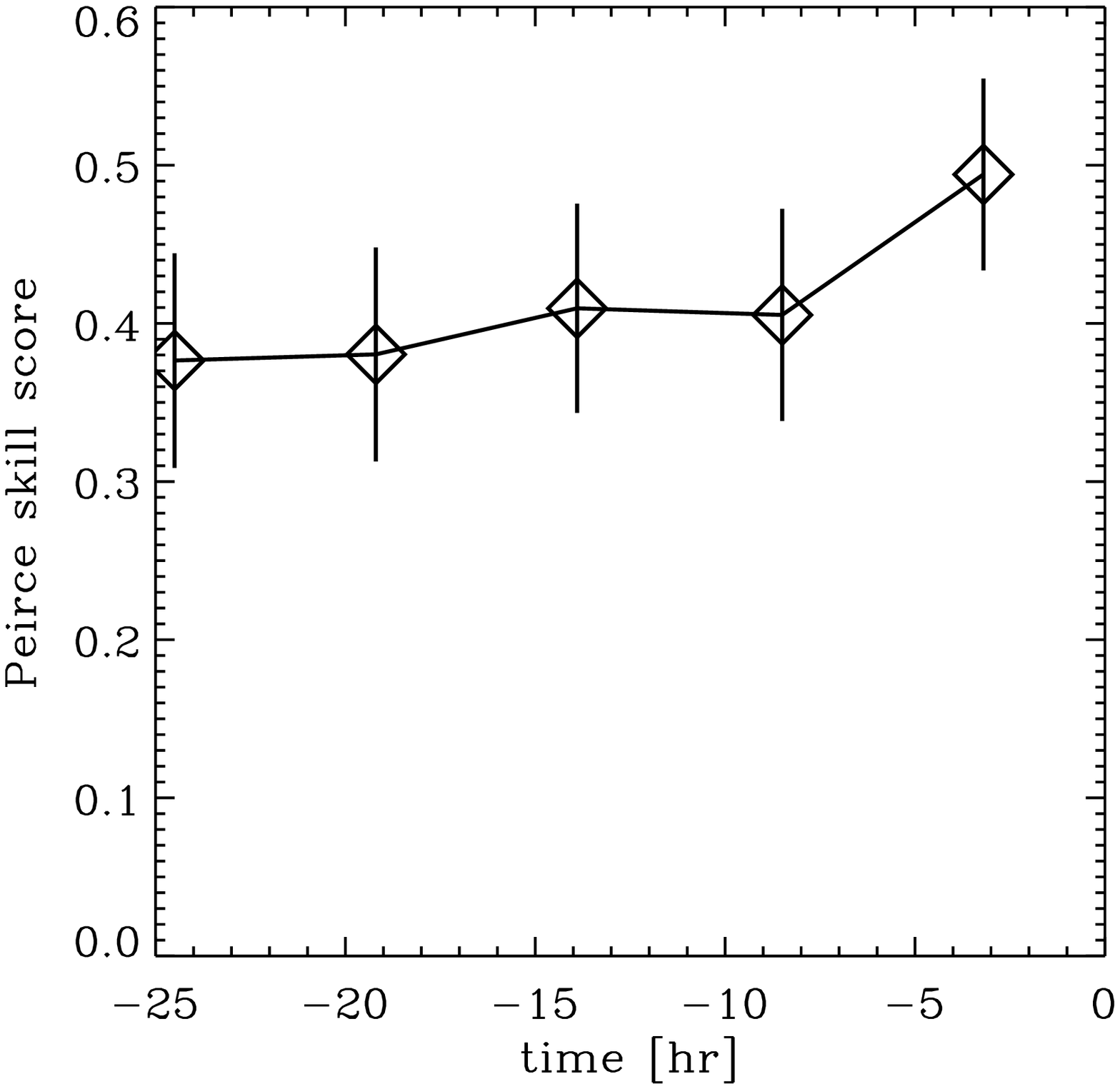}
\caption{Nonparametric discriminant analysis for the mean unsigned magnetic
flux, $\UFIELD$. 
{\it Left}: probability density estimates for the NE regions (black), and the
PE regions (red) in time interval 4, centered 3.2\,hr before the emergence
time, with the mean of each sample indicated by a vertical dashed line in the
corresponding color.  The shaded region is a 1-$\sigma$ estimate of the
uncertainty.  The discriminant boundary, where the two probability density
estimates are equal, is indicated by a vertical blue line; an observation to
the right of the boundary would be classified as an emergence.  There is an
obvious difference between the density estimates for the NE and PE regions: the
PE distribution has a longer tail to high field strength, but there is also
considerable overlap of the distributions.
{\it Right}: evolution of the Peirce skill score.  There is perhaps a weak
increase in the performance of $\UFIELD$ closer to the emergence time that is
likely due to the onset of emergence during time interval 4 for some regions,
but for most of the time intervals shown, the skill score is approximately
constant.
}
\label{fig:npda_field} 
\end{figure*}

The left panel of Figure~\ref{fig:npda_field} shows the probability density
estimates for the mean unsigned field strength, $\UFIELD$.  The peak of the PE
distribution is at a slightly higher field strength, and has a substantially
longer tail to large values, leading to a substantially higher mean value for
the PE sample than for the NE sample.  In this case, the large separation of
the means is misleading because of the presence of a few strong field PE
regions while the distributions of PE and NE for $\UFIELD$ show considerable
overlap.  The Peirce skill score for this variable is $0.49\pm0.06$.  The
discriminant boundary falls at $\UFIELD \approx 13$\,G; regions with a stronger
average unsigned field strength would be classified as PE. 

The right panel of Figure~\ref{fig:npda_field} shows the performance of
$\UFIELD$ as a function of time prior to emergence.  There is perhaps an
increase in the skill score from approximately one day before emergence to a
few hours before emergence, but the overall increase is not large in magnitude.
Certainly between 24.5\,hr and 8.5\,hr before emergence, the variations in the
skill score are less than the uncertainty; there is a small increase in the
last point, 3.2\,hr prior to emergence, which is likely to be a result of an
incorrect emergence time for some regions, so that surface field is appearing
during the final time interval (c.f.~Fig.~11 of \citetalias{Lekaetal2013}).  

\begin{figure*}
\plottwo{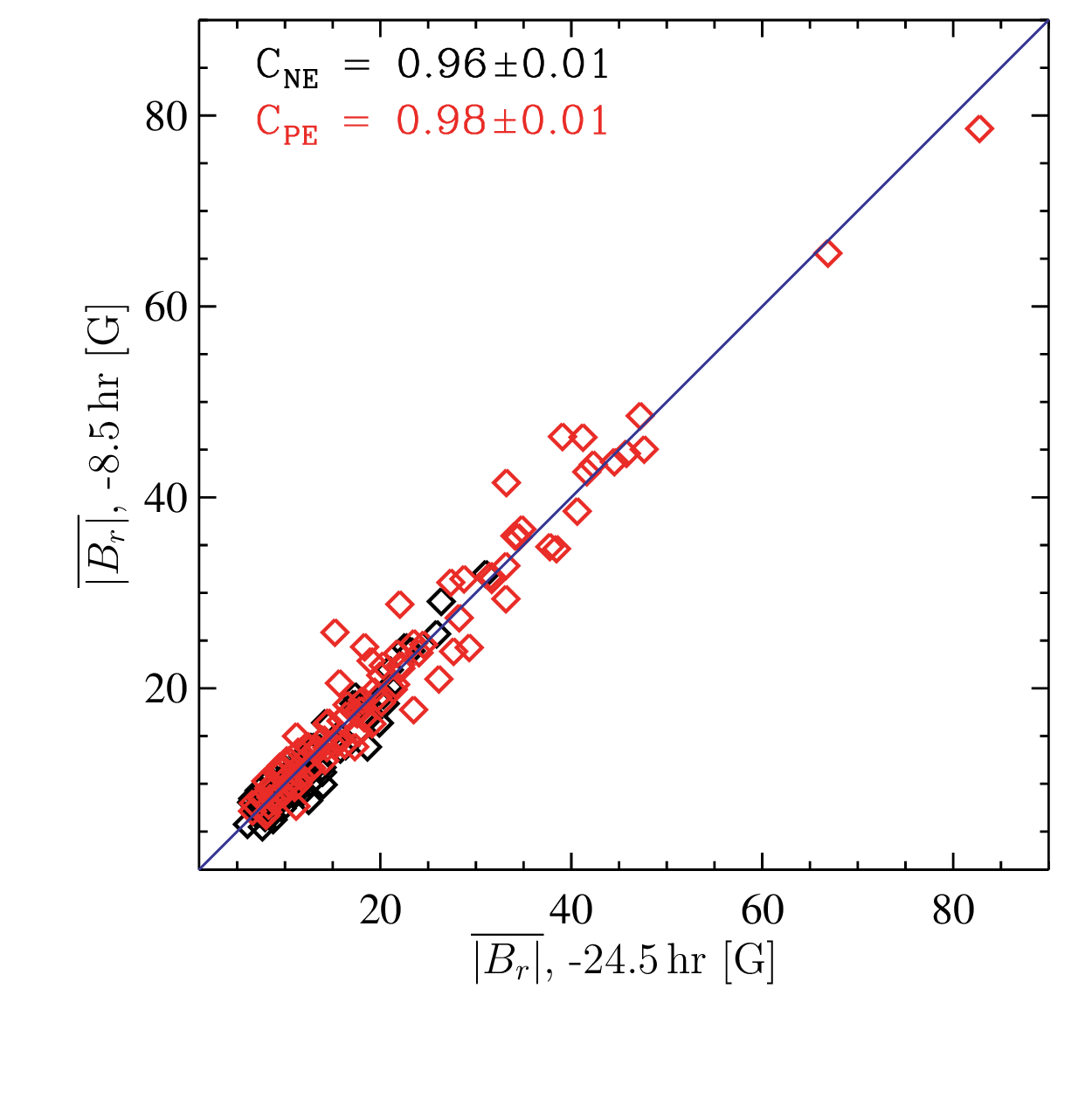}
{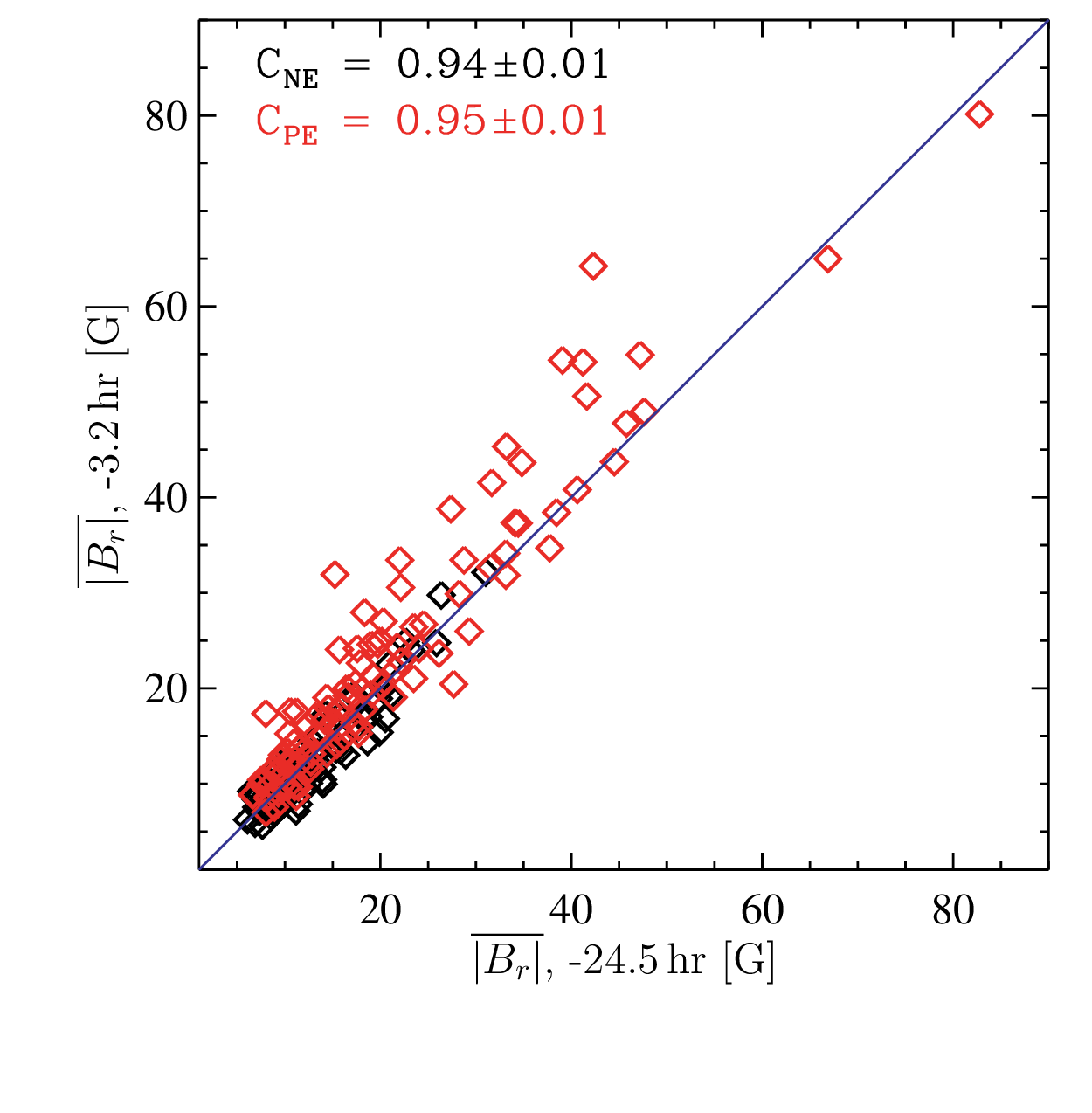}
\caption{{\it Left}: scatter plot of $\UFIELD$ for the time interval centered
24.5\,hr emergence versus the time interval centered 8.5\,hr before emergence
for NE (black) and PE (red) regions.  The Pearson correlation coefficient
between the PE at these times is $0.98 \pm 0.01$, while for the NE it is $0.96
\pm 0.01$.  
{\it Right}: same scatter plot but between 24.5\,hr and 3.2\,hr before
emergence.  The correlation coefficients are both slightly lower at $0.95 \pm
0.01$ for PE and $0.94 \pm 0.01$ for NE, suggesting that much of the evolution
of the flux occurs between 8.5\,hr and 3.2\,hr before emergence, which is
consistent with the emergence process beginning (for some regions) during the
final six hours before the time of emergence.
}
\label{fig:corr2_flux}
\end{figure*}

This constancy in the performance of the unsigned field is largely because the
field itself does not evolve substantially over the time in question.
Figure~\ref{fig:corr2_flux} (left) shows the field 24.5\,hr before emergence
versus the field 8.5\,hr before emergence.  There is an extremely high
correlation between the field at the two times (Pearson correlation
coefficients $\ge 0.96$) for both the PE and NE regions,  and no clear
indication of evolution.  When considering the change between 24.5\,hr and
3.2\,hr before emergence (Fig.~\ref{fig:corr2_flux}, right), there is some
indication of evolution of the field, consistent with there being a few regions
for which emergence began in the final six hours before the nominal emergence
time (more points lying above and to the left of the blue line).  This trend is
still weak compared to the variation among the regions considered, as is born
out by the small decrease in the correlations (Pearson correlation coefficients
$\ge 0.94$).  Thus the evolution of the field does not greatly change the
ability of the unsigned field to distinguish the PE from the NE.  

The large overlap between the distributions of NE and PE regions shows that
there is no clear signature in the surface field when individual regions are
considered (c.f.~Fig.~2 and 3 of \citetalias{Birchetal2013}).  However, there
was a bias introduced in the selection criteria for the NE compared with the
PE: NE regions were required to have magnetic field consistently $<1000$\,G
(see \S3.2 of \citetalias{Lekaetal2013}), while no such requirement was imposed
for PE regions.  It may simply be that the difference between the PE and NE
samples results from the bias in the selection of NE compared to the PE
regions.

It is also possible that there is a small amount of weak magnetic flux present
at the surface more than a day prior to the beginning of the clear emergence
phase of active regions.  This field is indistinguishable from noise in
individual MDI magnetograms, but becomes apparent in averaging over large
numbers.  This could be related to the emergence process, in the form of small
amounts of flux arriving at the emergence site prior to the main emergence, as
is seen in some simulations \citep[e.g.,][]{Cheungetal2010,Steinetal2011}.  It
could also be related to the known tendency for active regions to emerge in the
same locations as prior active regions \citep[e.g.,][]{PojogaCudnik2002}.  The
latter case would be one example of how the bias manifests from a completely
solar cause.

\subsection{Measures of the Center-to-Annulus and Annulus-to-Center Travel
Times}

\begin{figure*}
\plottwo{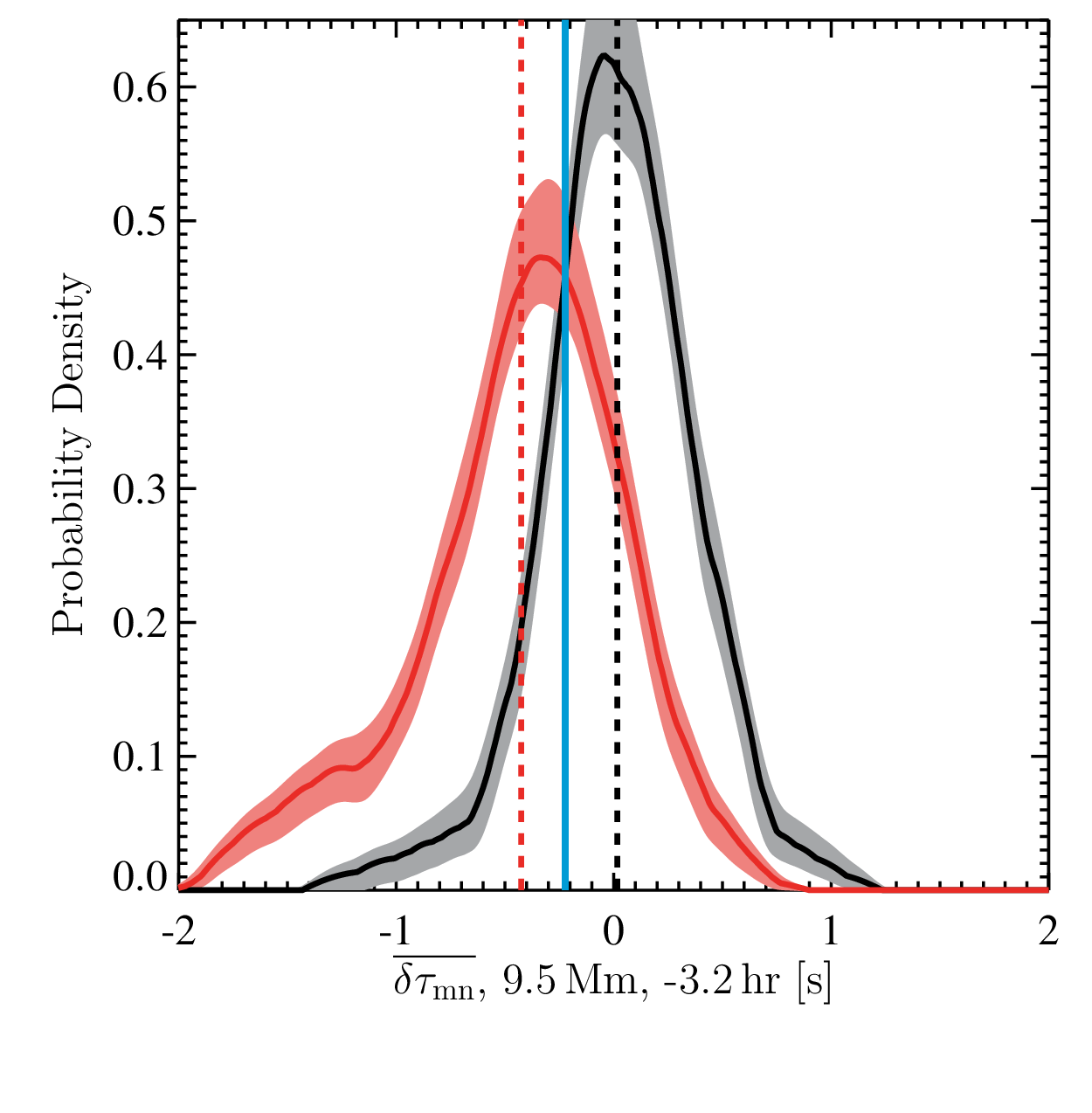}
{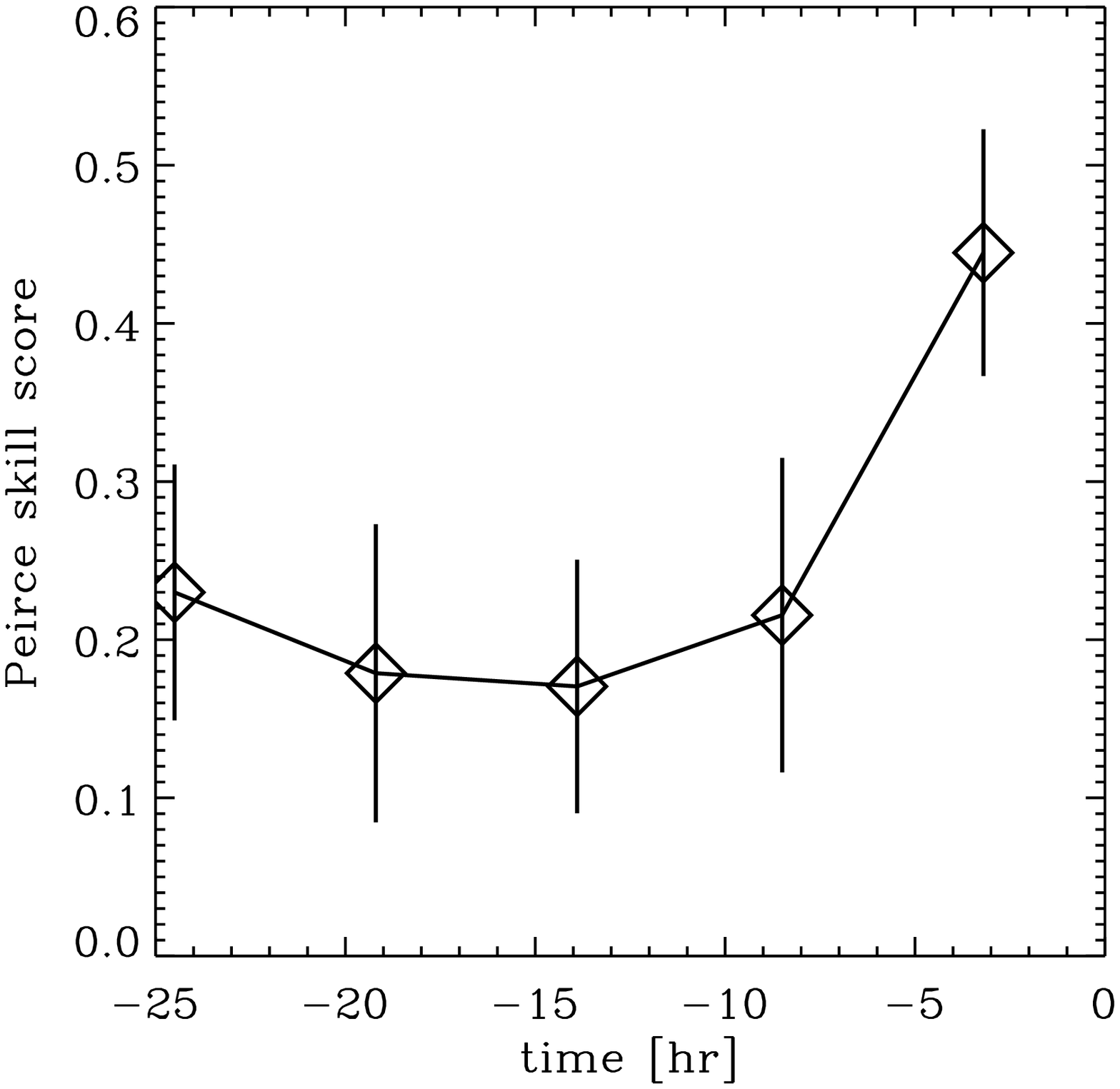}
\caption{Nonparametric discriminant analysis for the mean travel-time
shift, $\DTMEAN$, in filter TD5, in the same format as
Fig.~\ref{fig:npda_field}. 
{\it Left}: there is an obvious difference between the density estimates for
the NE and PE regions in the time interval centered 3.2\,hr before the
emergence time, with the PE regions typically having negative mean
travel-time shifts.
{\it Right}: the performance of $\DTMEAN$ shows a similar trend to $\UFIELD$,
constant for most of the time considered, with an increase at the last time
interval, although the skill score is consistently lower for $\DTMEAN$.
}
\label{fig:npda_dtmean} 
\end{figure*}

Immediately prior to emergence, the mean travel-time shift measured in a
variety of filters shows a significant ability to distinguish PE from NE
regions.  The left panel of Figure~\ref{fig:npda_dtmean} shows the mean
travel-time, $\DTMEAN$, in filter TD5, for time interval 4 (centered 3.2\,hr
prior to the time of emergence).  The NE sample has a mean very close to 0\,s,
and its distribution is symmetric and peaked close to 0\,s, consistent with the
differences from 0\,s being simply due to noise.  The mean of the PE sample is
negative and close to the peak in its distribution; the PE distribution is
slightly wider than the NE distribution.  Although there is a distinct
difference visible in the distributions, there is also substantial overlap
between the two, as in the $\UFIELD$ case.  This is quantified by a Peirce
skill score of $0.45\pm0.08$.  Regions with a mean travel-time shift less than
the discriminant boundary (at approximately -0.2\,s) would be classified as PE,
while those above would be classified as NE.  That is, negative mean
travel-time shifts are associated with emerging regions.  Like the average
unsigned field, the performance of the mean travel-time shift is better 3.2\,hr
before emergence than at earlier times (compare Fig.~\ref{fig:npda_field},
right to Fig.~\ref{fig:npda_dtmean}, right). 

\begin{figure}
\plotone{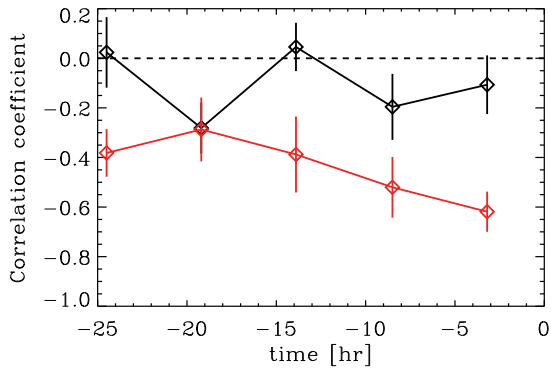}
\caption{Pearson correlation coefficient between $\UFIELD$ and $\DTMEAN$ in
filter TD5 from a bootstrap method for the NE (black) and PE (red) regions as a
function of time.  For the NE regions, the correlation coefficient is generally
close to zero, as expected.  For the PE regions, there is a weak trend towards
stronger (negative) correlations at later times. 
} 
\label{fig:corr2_segment}
\end{figure}

The mean travel-time shift is typically reduced in the presence of surface magnetic
field \citep[e.g.,][]{LindseyBraun2005,BraunBirch2008}.
Figure~\ref{fig:corr2_segment} shows the Pearson correlation coefficient
between $\DTMEAN$ and $\UFIELD$ as a function of time.  The correlation
coefficient for the NE regions is generally close to 0, as would be expected if
the mean travel-time shifts are simply due to noise, while for the PE regions,
the correlation coefficient is negative, with perhaps a weak trend towards a
stronger (negative) correlation closer to the emergence time, although there is
not a distinct difference between the final time interval prior to emergence
and earlier time intervals.  It is possible that the difference between the PE
and NE regions in $\DTMEAN$ is simply an indirect result of the difference in
the surface field.  However, it is also possible that there is a signal in
$\DTMEAN$ during all the time intervals that is not a result of the surface
magnetic field.  The influence of the surface field on the travel-times is
investigated further in \S\ref{sec:fluxmatch}. 

There are also many instances where $\DTIN$ and $\DTOUT$ have a skill score
only slightly less than $\DTMEAN$ in the same filter.  In all time intervals
and filters, there is a moderate correlation between $\DTIN$ and $\DTOUT$.
Since $\DTMEAN$ is a linear combination of $\DTIN$ and $\DTOUT$, it is likely
that the slightly better performance of $\DTMEAN$ is simply a result of a
better signal to noise ratio than either $\DTIN$ or $\DTOUT$ considered alone. 

\subsection{East-West and North-South Travel Times}

\begin{figure*}
\plottwo{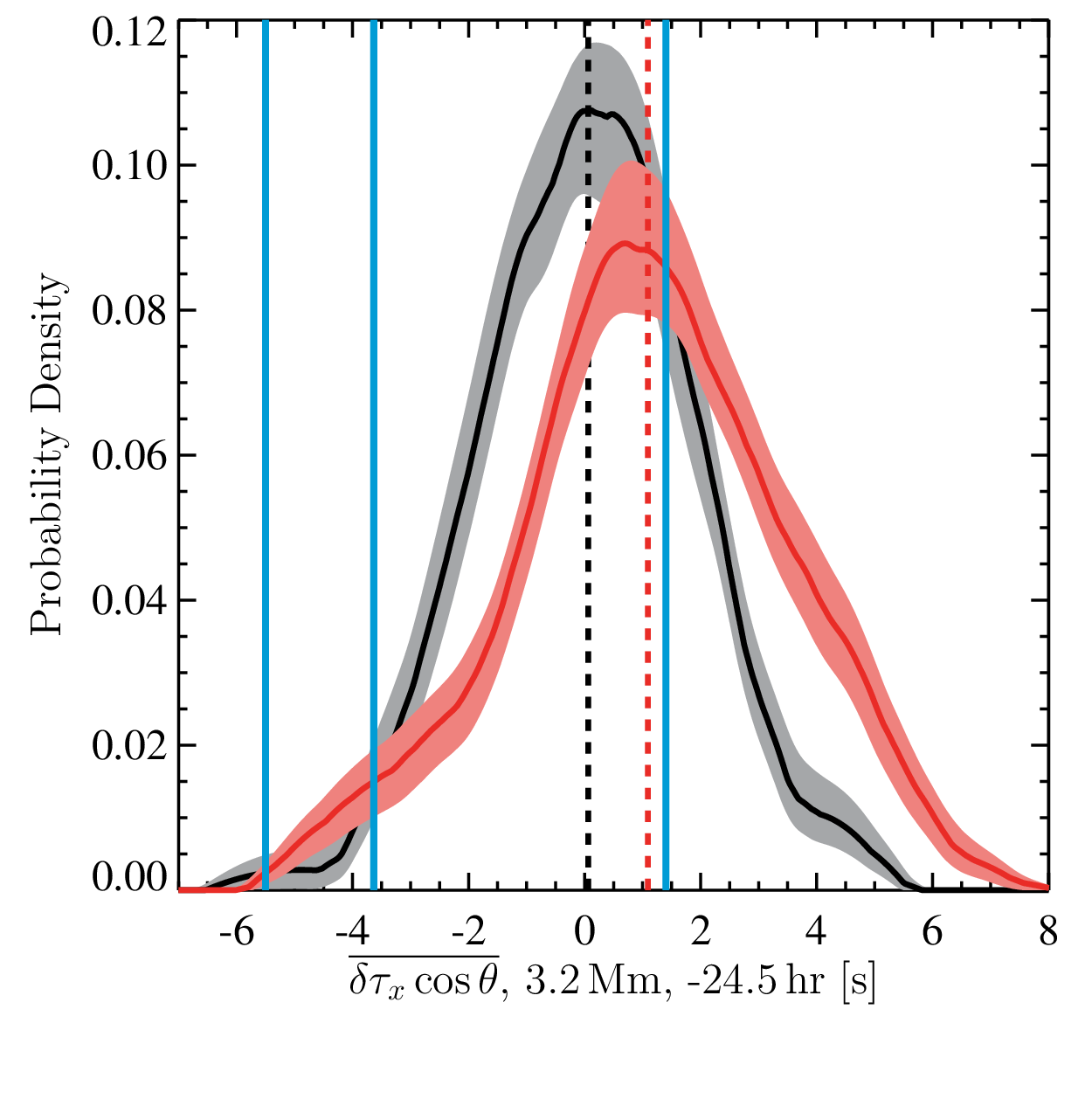}
{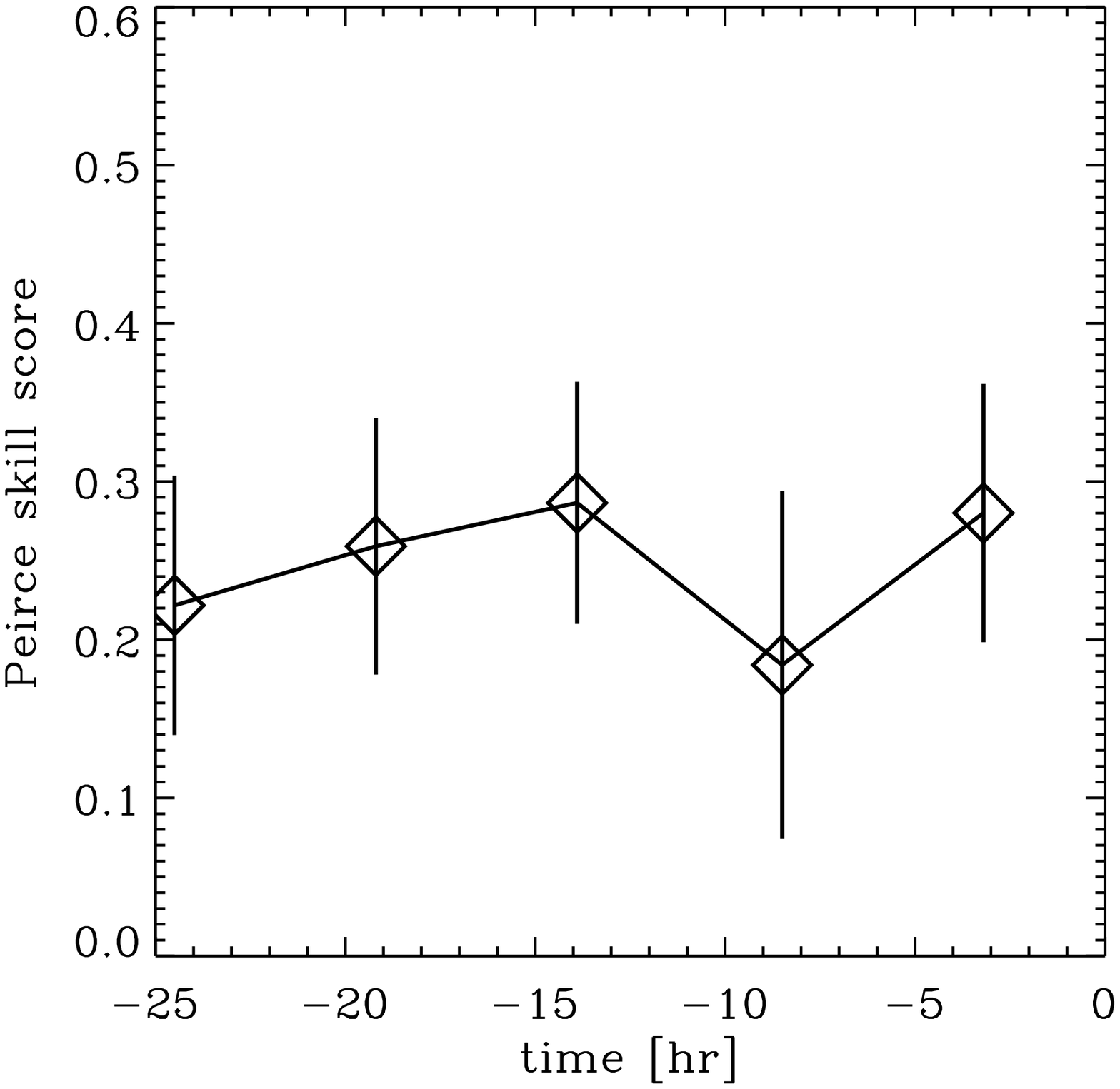}

\plottwo{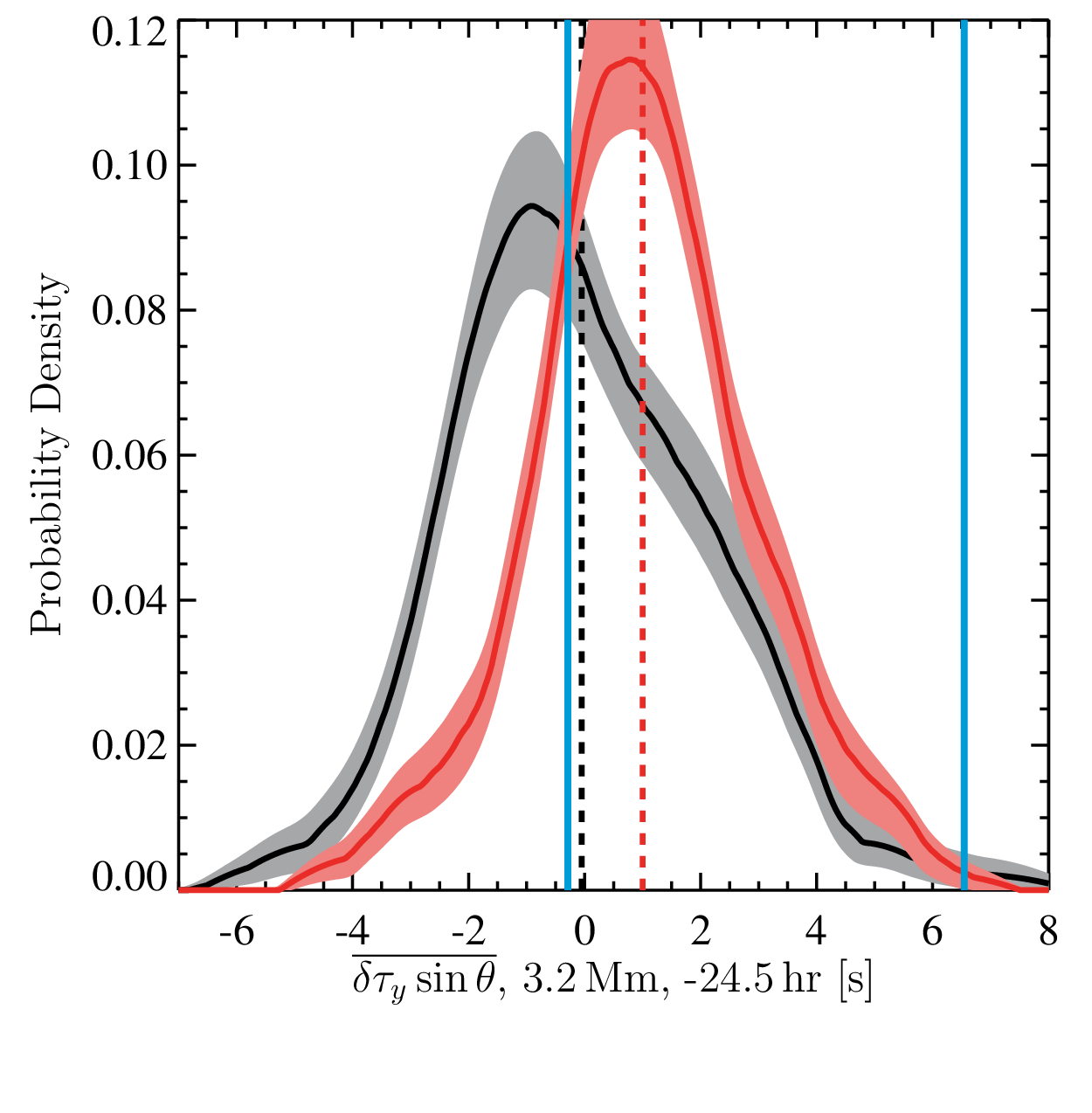}
{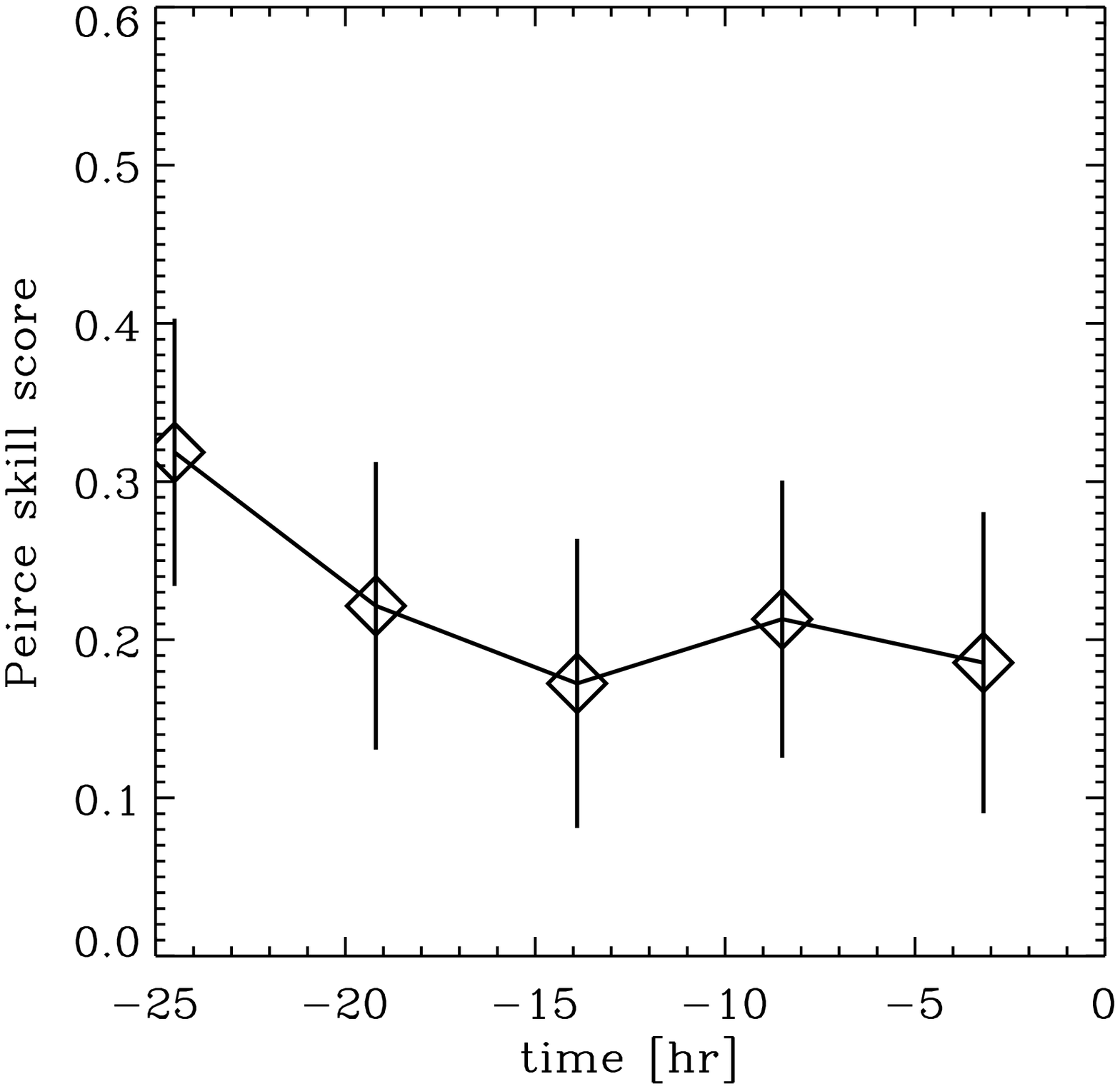}
\caption{Nonparametric discriminant analysis for antisymmetric averages of the
east-west and north-south travel-time differences, $\DTXC$ (top) and $\DTYS$
(bottom) in filter TD3, in the same format as Fig.~\ref{fig:npda_field}.  
{\it Left}: the density estimates for NE regions for both $\DTXC$ and $\DTYS$,
24.5\,hr before the emergence time, are peaked close to 0\,s, and are fairly
symmetric; the density estimates for PE regions are both peaked close to 1\,s.
{\it Right}: given the uncertainties, there is no obvious trend in $\DTXC$ and
only weak evidence for a trend towards worse performance closer to emergence
for $\DTYS$.
} 
\label{fig:npda_dtxy}
\end{figure*}

In Table~\ref{tbl:peirce_boot_npda_equal_n1_seg}, antisymmetric averages of the
east-west and north-south travel-time differences appear most frequently at shallow to
moderate depths and earlier time intervals.  The left panels of
Figure~\ref{fig:npda_dtxy} show the distributions of $\DTXC$ and $\DTYS$ in
filter TD3, centered 24.5\,hr before emergence.  For both variables, the mean
and the peak of the NE distributions lie close to 0\,s, while the mean and the
peak of the PE distributions are at positive travel-time differences of
approximately 1\,s.  Unlike previous variables considered, there are multiple
discriminant boundaries; the boundaries in the tails of the distributions, at
large negative values of $\DTXC$ and large positive values of $\DTYS$, are
likely to be spurious results caused by a few regions having extreme values. 

The evolution of the skill score (Fig.~\ref{fig:npda_dtxy}, right) shows that
the variations with time are unlikely to be real given the uncertainties in the
resulting skill scores, although there is perhaps a trend for worse performance
of $\DTYS$ at times closer to the emergence time.  However, this is one example
of a variable that, in one time interval, falls above the threshold to be
included in Table~\ref{tbl:peirce_boot_npda_equal_n1_seg}, while in other time
intervals, it may be excluded from the table, despite have substantial ability
to discriminate PE from NE regions.

The particular combinations that appear, namely $\DTXC$ and $\DTYS$, would be
expected to have a signal from a converging flow. However, note that only one
instance of $\DTOI$ appears in this table, and in a filter with a much deeper
lower turning point.  The relative strength of the signals in $\DTXC$ and
$\DTYS$ versus the signal in $\DTOI$ in general depends on the geometry of the
assumed flow; detailed modeling is beyond the scope of the current work.  We
note, however, that Figure~5 of \citetalias{Birchetal2013} shows patterns in
the ensemble averages of $\DTXC$ and $\DTYS$ with more structure than would be
expected for a simple converging flow.  As discussed in
\citetalias{Birchetal2013}, one potential interpretation for the signals in
$\DTXC$ and $\DTYS$ is a preference for emergence to occur at the boundary
between supergranules, so these signals are the result of supergranular flows,
not the emergence process itself. 

\subsection{The Vorticity}

\begin{figure*}
\plottwo{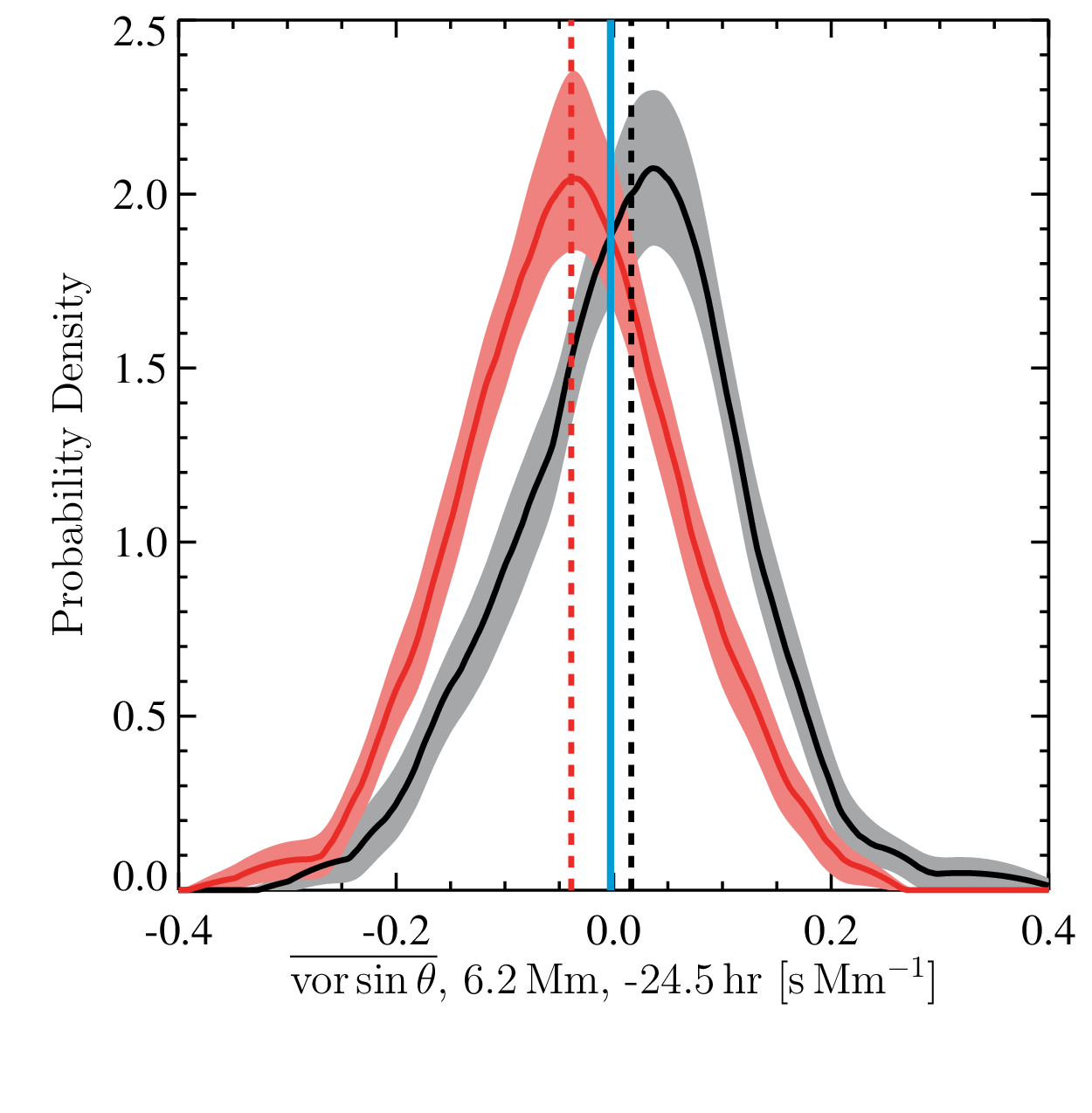}
{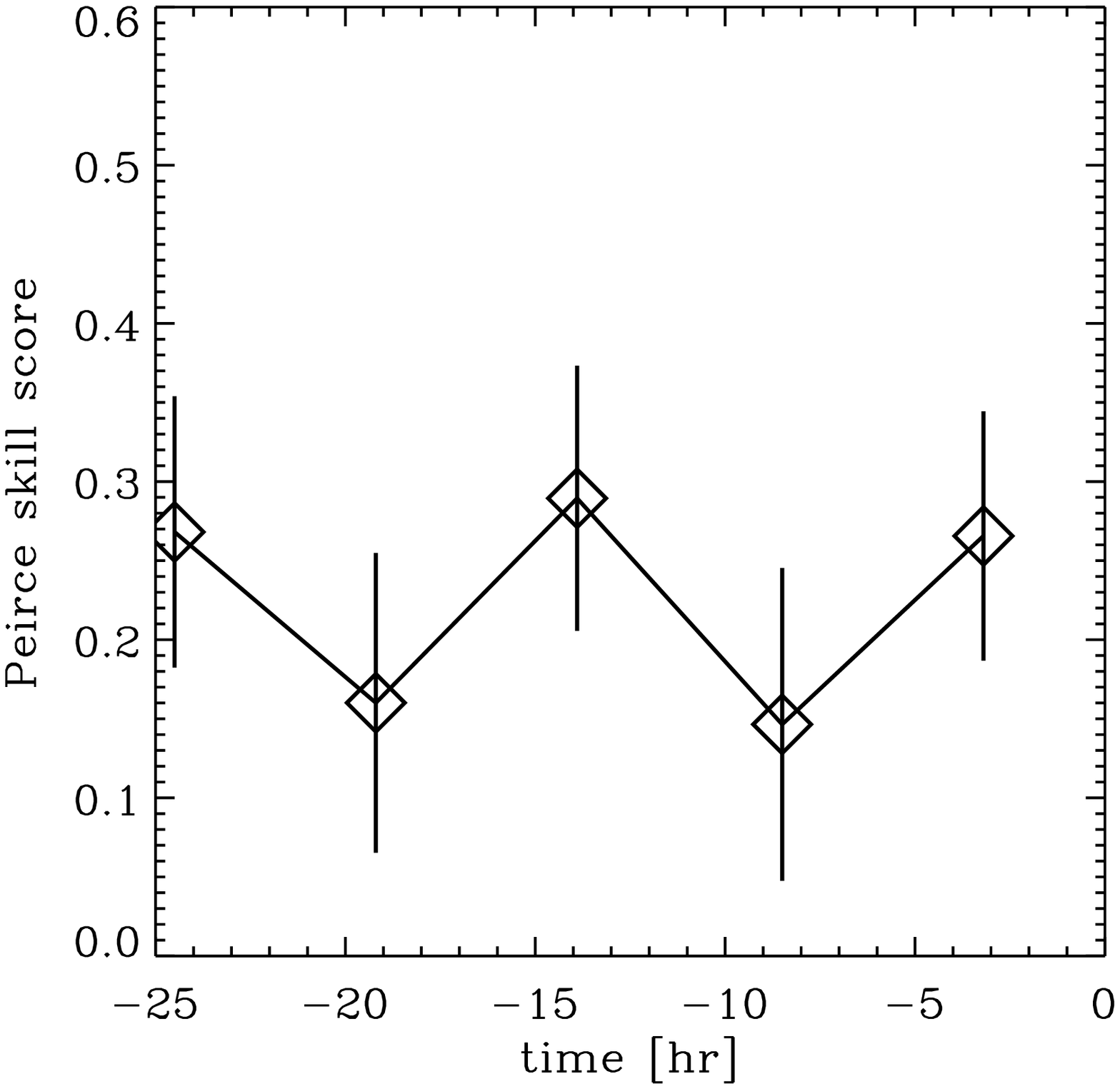}
\caption{Nonparametric discriminant analysis for an antisymmetric average of
the vertical vorticity, $\VORS$, in filter TD4, in the same format as
Fig.~\ref{fig:npda_field}.  
{\it Left}: 24.5\,hr before the emergence time, the density estimate for NE
regions has a peak at and a mean value that are slightly positive, while the
peak and mean value for the PE distribution are at negative values. 
{\it Right}: given the uncertainties in the skill score, the performance of
this variable does not clearly evolve with time.
} 
\label{fig:npda_vors}
\end{figure*}

The remaining variable appearing multiple times in the list of best parameters
is $\VORS$, particularly in filter TD4.  Figure~\ref{fig:npda_vors} (left)
shows a small but clear offset in the distributions of the NE and PE regions
24.5\,hr before the emergence time, with PE regions more likely to have
negative values of $\VORS$ relative to NE regions.  As for the variables
$\DTXC$ and $\DTYS$, this is an example of a variable that, in some time
intervals, falls above the threshold to be included in
Table~\ref{tbl:peirce_boot_npda_equal_n1_seg}, while in other time intervals,
it is excluded from the table, despite have substantial ability to discriminate
PE from NE regions (see Fig.~\ref{fig:npda_vors}, right).

It is known that surface magnetic fields are associated with a prograde flow
\citep[e.g.,][]{Zhaoetal2004}.  Such a prograde flow would result in the signal
seen in $\VORS$, so one explanation for this is that, once again, the
difference between the PE and NE regions is a result of the surface magnetic
field.  To determine whether there is a helioseismic signature of the emergence
process not caused by the surface magnetic field, it is important to account
for the contribution of the surface field to the differences in the
travel-times.

\subsection{Matching the Flux Distribution}\label{sec:fluxmatch}

To investigate the influence of the surface magnetic field on the helioseismic
parameters, we selected subsets of the NE and PE regions with matching
distributions of average unsigned field, $\UFIELD$, and location on the disk.
The approach to this was essentially the same as the selection of the NE
regions to match the distributions in position and time of the PE regions
described in \citetalias{Lekaetal2013}: we used simulated annealing to select
subsets of PE and NE regions that minimize the integrated absolute value of the
difference between nonparametric density estimates for the two distributions.
Sixty-five regions from each sample were selected, which resulted in 50--55
regions with good duty cycle in each sample.  This was the largest subset for
which an integrated absolute difference of no more than about 0.1 could be
obtained; the integrated absolute difference of two completely non-overlapping
distributions would be 2.

We performed the same analysis as for the full set of regions on these subsets
of regions to rank the variables by skill score.  We also repeated the Monte
Carlo experiment (see appendix~\ref{sec:MC}) for these sample sizes.  We found
that although the maximum skill scores of the helioseismology variables can
reasonably be expected when there is no difference in the populations, there is
still a preponderance of large skill score values in the helioseismology
variables compared to what would be expected from chance. This suggests that
some of the helioseismology variables have real ability to discriminate the PE
regions from the NE regions, but that it is difficult to determine if any
specific variable has any power to discriminate.  Because of this, we choose to
show variables with Peirce skill scores greater than 0.24 (rather than the 0.27
used for the full samples) in
Table~\ref{tbl:subsetpeirce_boot_npda_equal_n1_seg}.  This is done to highlight
the variables (shaded in the table) that still have large skill scores after
the flux matching.  

\begin{deluxetable}{crc}
\tablecolumns{3}
\tablewidth{0pc}
\tablecaption{Best Performing Variables\label{tbl:subsetpeirce_boot_npda_equal_n1_seg}}
\tablehead{
\colhead{variable} & \colhead{depth} & \colhead{Peirce SS} \\
\colhead{} & \colhead{(Mm)} & \colhead{}}
\startdata
\cutinhead{TI-0: time=$t_0-24.5$\,hr}
$\overline{\delta \tau_y \sin \theta}^\dagger$ &  3.2 & $ 0.33\pm 0.10$ \\
$\overline{\delta \tau_y \sin \theta}^\dagger$ &  2.2 & $ 0.32\pm 0.11$ \\
$\overline{\VOR \sin \theta}^\dagger$ & 15.7 & $ 0.31\pm 0.12$ \\
$\overline{\delta \tau_x}$ & 20.9 & $ 0.29\pm 0.10$ \\
$\overline{\DIV}$ &  9.5 & $ 0.27\pm 0.10$ \\
$\overline{\VOR \sin \theta}^\dagger$ &  6.2 & $ 0.27\pm 0.11$ \\
$\overline{\delta \tau_x \cos \theta}$ &  6.2 & $ 0.24\pm 0.11$ \\
$\overline{\delta \tau_{\rm io} \cos \theta}$ &  3.2 & $ 0.24\pm 0.11$ \\
\cutinhead{TI-1: time=$t_0-19.2$\,hr}
$\overline{\delta \tau_{\rm out} \cos \theta}$ &  9.5 & $ 0.25\pm 0.10$ \\
$\overline{\DIV \sin \theta}$ & 15.7 & $ 0.25\pm 0.12$ \\
$\overline{\delta \tau_{\rm in} \cos \theta}$ &  6.2 & $ 0.25\pm 0.11$ \\
$\overline{\VOR \sin \theta}$ & 11.4 & $ 0.25\pm 0.11$ \\
$\overline{\delta \tau_x}$ &  2.2 & $ 0.24\pm 0.11$ \\
$\overline{\delta \tau_x \cos \theta}$ &  9.5 & $ 0.24\pm 0.12$ \\
$\overline{\delta \tau_{\rm in} \cos \theta}$ & 15.7 & $ 0.24\pm 0.11$ \\
$\overline{\delta \tau_{\rm io} \cos \theta}$ &  9.5 & $ 0.24\pm 0.10$ \\
$\overline{\delta \tau_{\rm out}}$ &  3.2 & $ 0.24\pm 0.12$ \\
$\overline{\delta \tau_x}$ &  3.2 & $ 0.24\pm 0.11$ \\
\cutinhead{TI-2: time=$t_0-13.9$\,hr}
$\overline{\DIV \cos \theta}$ & 15.7 & $ 0.36\pm 0.10$ \\
$\overline{\VOR}$ & 23.3 & $ 0.30\pm 0.10$ \\
$\overline{\VOR \sin \theta}^\dagger$ &  6.2 & $ 0.29\pm 0.11$ \\
$\overline{\delta \tau_x \cos \theta}$ &  9.5 & $ 0.26\pm 0.11$ \\
$\overline{\DIV}$ &  6.2 & $ 0.26\pm 0.11$ \\
$\overline{\delta \tau_x \cos \theta}^\dagger$ &  3.2 & $ 0.25\pm 0.12$ \\
$\overline{\delta \tau_x}$ & 15.7 & $ 0.25\pm 0.13$ \\
$\overline{\DIV}$ & 13.3 & $ 0.24\pm 0.11$ \\
$\overline{\delta \tau_y \cos \theta}$ & 11.4 & $ 0.24\pm 0.12$ \\
$\overline{\delta \tau_x \cos \theta}$ & 15.7 & $ 0.24\pm 0.12$ \\
\cutinhead{TI-3: time= $t_0-8.5$\,hr}
$\overline{\DIV}$ & 11.4 & $ 0.36\pm 0.11$ \\
$\overline{\delta \tau_x \cos \theta}$ &  2.2 & $ 0.26\pm 0.12$ \\
$\overline{\VOR \cos \theta}$ &  6.2 & $ 0.25\pm 0.11$ \\
$\overline{\delta \tau_x \cos \theta}$ &  6.2 & $ 0.25\pm 0.12$ \\
$\overline{\delta \tau_{\rm io} \cos \theta}$ &  1.4 & $ 0.25\pm 0.12$ \\
$\overline{\delta \tau_x \cos \theta}$ &  3.2 & $ 0.25\pm 0.13$ \\
\cutinhead{TI-4: time= $t_0-3.2$\,hr}
$\overline{\delta \tau_{\rm mn}}^\dagger$ &  9.5 & $ 0.38\pm 0.09$ \\
$\overline{\VOR \sin \theta}^\dagger$ &  6.2 & $ 0.33\pm 0.11$ \\
$\overline{\delta \tau_{\rm out}}^\dagger$ &  9.5 & $ 0.30\pm 0.10$ \\
$\overline{\delta \tau_x}$ &  6.2 & $ 0.29\pm 0.11$ \\
$\overline{\delta \tau_{\rm out} \cos \theta}$ & 13.3 & $ 0.29\pm 0.12$ \\
$\overline{\delta \tau_{\rm in}}^\dagger$ &  9.5 & $ 0.28\pm 0.12$ \\
$\overline{\delta \tau_{\rm mn}}^\dagger$ & 23.3 & $ 0.28\pm 0.12$ \\
$\overline{\VOR}$ &  2.2 & $ 0.28\pm 0.11$ \\
$\overline{\delta \tau_{\rm out} \cos \theta}$ &  9.5 & $ 0.27\pm 0.11$ \\
$\overline{\VOR \cos \theta}$ & 23.3 & $ 0.26\pm 0.11$ \\
$\overline{\delta \tau_{\rm io} \cos \theta}$ & 20.9 & $ 0.25\pm 0.11$ \\
$\overline{\delta \tau_{\rm out} \cos \theta}$ & 11.4 & $ 0.24\pm 0.11$ \\
$\overline{\delta \tau_{\rm mn}}^\dagger$ & 20.9 & $ 0.24\pm 0.11$ \\
$\overline{\delta \tau_x}$ &  3.2 & $ 0.24\pm 0.13$ \\
$\overline{\delta \tau_{\rm mn}}$ & 18.2 & $ 0.24\pm 0.10$ \\
$\overline{\delta \tau_x \cos \theta}^\dagger$ &  3.2 & $ 0.24\pm 0.12$ \\
$\overline{\delta \tau_y \sin \theta}^\dagger$ &  2.2 & $ 0.24\pm 0.13$ \\
\enddata
\tablecomments{Depth refers to the lower turning point of the waves in the
filter used.  Time is relative to the emergence time $t_0$.  Variables marked with a ${}^\dagger$
also appear in Table~\ref{tbl:peirce_boot_npda_equal_n1_seg}. A version of
Table~\ref{tbl:subsetpeirce_boot_npda_equal_n1_seg} containing all the
variables considered is published in the electronic edition of Barnes et
al.~(2013).}
\end{deluxetable}

\begin{figure*}
\plottwo{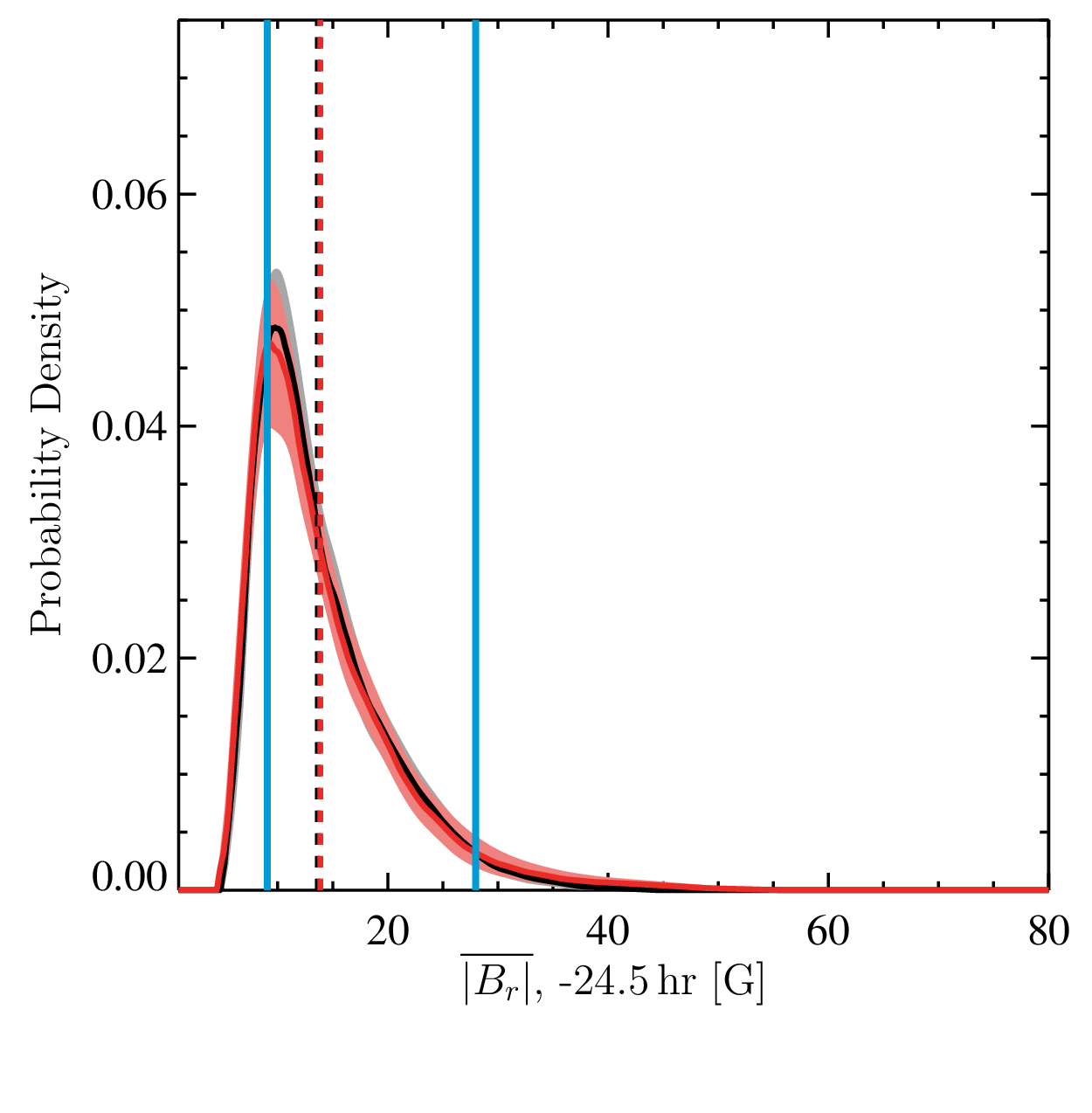}
{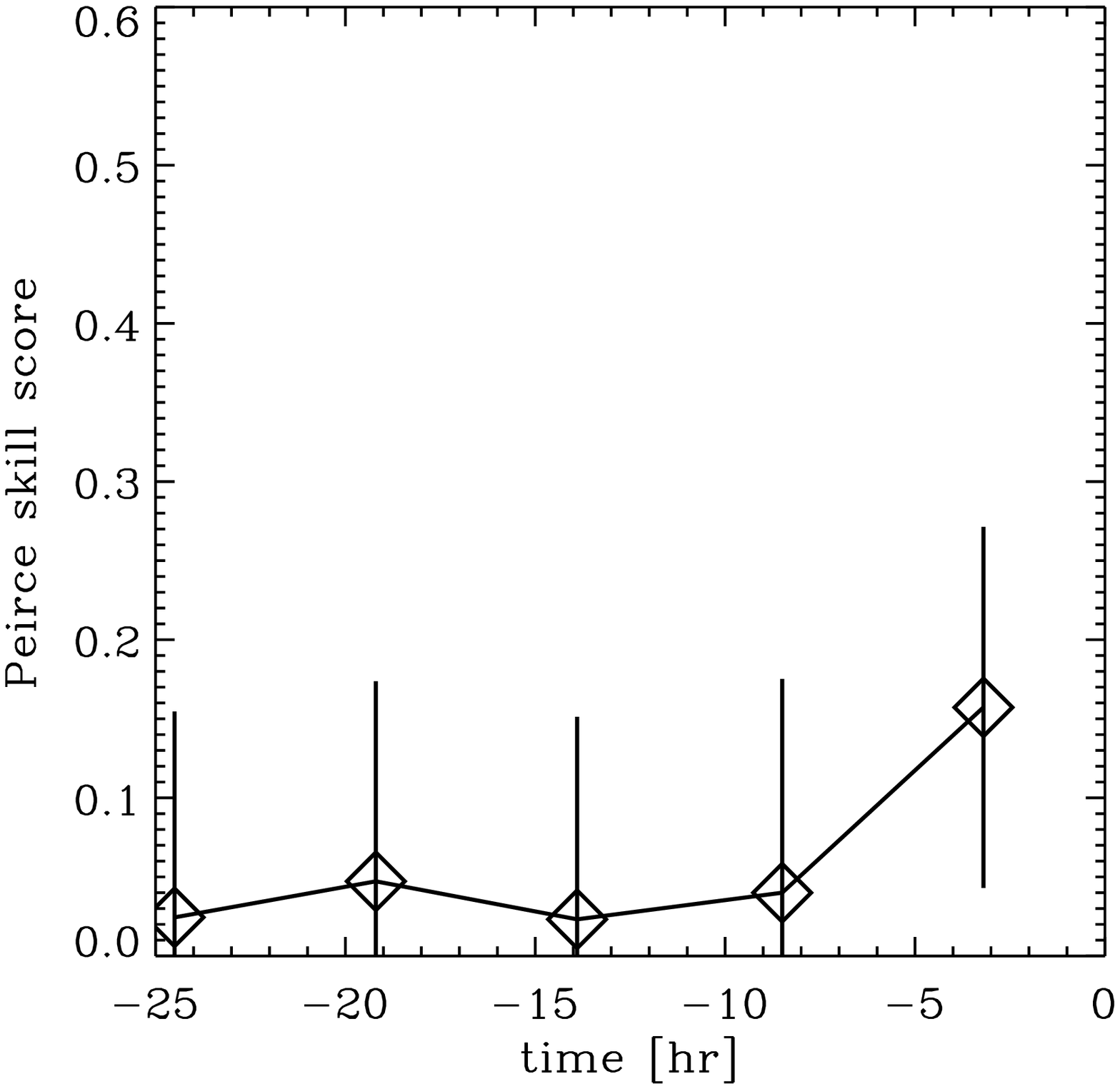}
\caption{Nonparametric discriminant analysis for the mean unsigned magnetic
flux, $\UFIELD$, for a subset of regions with matched distributions of magnetic
flux, in the same format as Fig.~\ref{fig:npda_field}.  By construction of the
subset, the distributions of PE and NE regions are closely matched, as seen in
the left panel, thus the Peirce skill score is consistent with 0 for time
intervals 0--3, as it should be for matched distributions, with a slight
increase at the final time interval likely due to the onset of emergence during
time interval 4 for some regions.
}
\label{fig:npda_field_subset} 
\end{figure*}

Due to the matching of the distributions, $\UFIELD$ has virtually no ability to
discriminate between the samples, and thus does not appear anywhere in
Table~\ref{tbl:subsetpeirce_boot_npda_equal_n1_seg}.  To illustrate how well
the distributions match, the distributions of $\UFIELD$ in time interval 0, a
day before emergence, are shown in Figure~\ref{fig:npda_field_subset}, along
with the skill score as a function of time.  The skill score is consistent with
zero in time intervals 0--3, and compared with Figure~\ref{fig:npda_field}, the
PE and NE distributions are very closely matched, with no remaining tail to
large values of $\UFIELD$ for the PE sample.  The increase in skill score in
time interval 4 is likely due to the start of emergence in a few regions.

For most of the variables that are not strongly correlated with the magnetic
flux, the skill score values have not changed substantially, but the smaller
sample sizes generally lead to larger uncertainties.  Almost every filter and
depth combination of $\DTXC$, $\DTYS$ and $\VORS$ present in
Table~\ref{tbl:peirce_boot_npda_equal_n1_seg} is also present after flux matching in
Table~\ref{tbl:subsetpeirce_boot_npda_equal_n1_seg}, with a similar value of
the skill score.  For example, in time interval 0, the skill score values
$\DTYS$ and $\VORS$ for the filters found in
Table~\ref{tbl:peirce_boot_npda_equal_n1_seg} lie in the range $0.27 \le {\rm
PSS} \le 0.34$ for both the full samples and the matched flux subsets.  By
contrast, the mean travel-time shifts in time interval 4 have consistently
lower skill scores for the matched flux subset, indicating that the correlation
with the magnetic flux accounts for some of the ability of $\DTMEAN$ (and
$\DTIN$ and $\DTOUT$) to discriminate between PE and NE regions.  In addition
to the best performing variables for the full samples of PE and NE regions,
there are a considerable number of other variables, in a range of filters,
present in Table~\ref{tbl:subsetpeirce_boot_npda_equal_n1_seg}.  Many of these
are likely to be statistical anomalies, with no real ability to discriminate
the PE and NE regions. 

\begin{figure*}
\plottwo{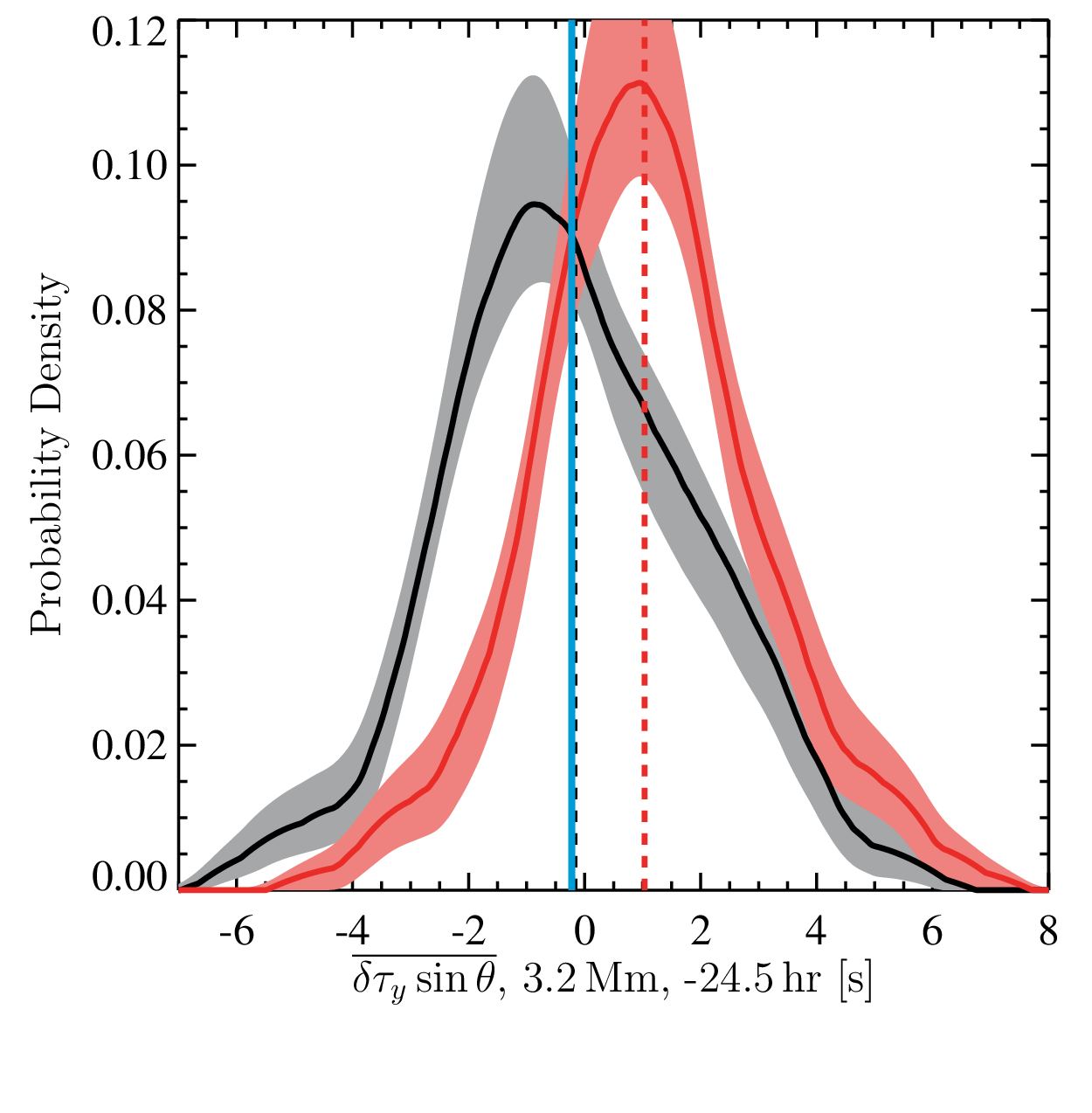}
{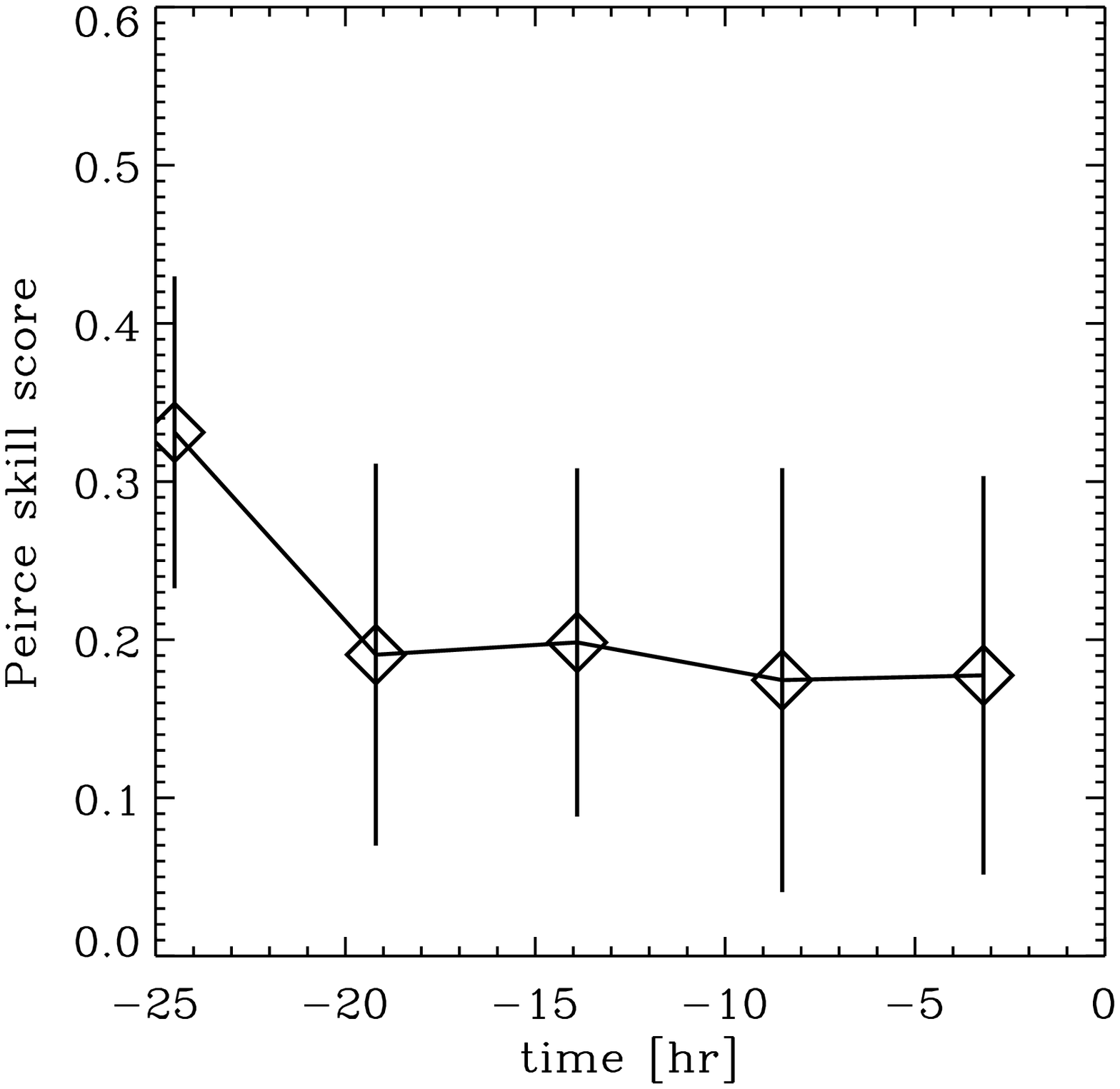}

\plottwo{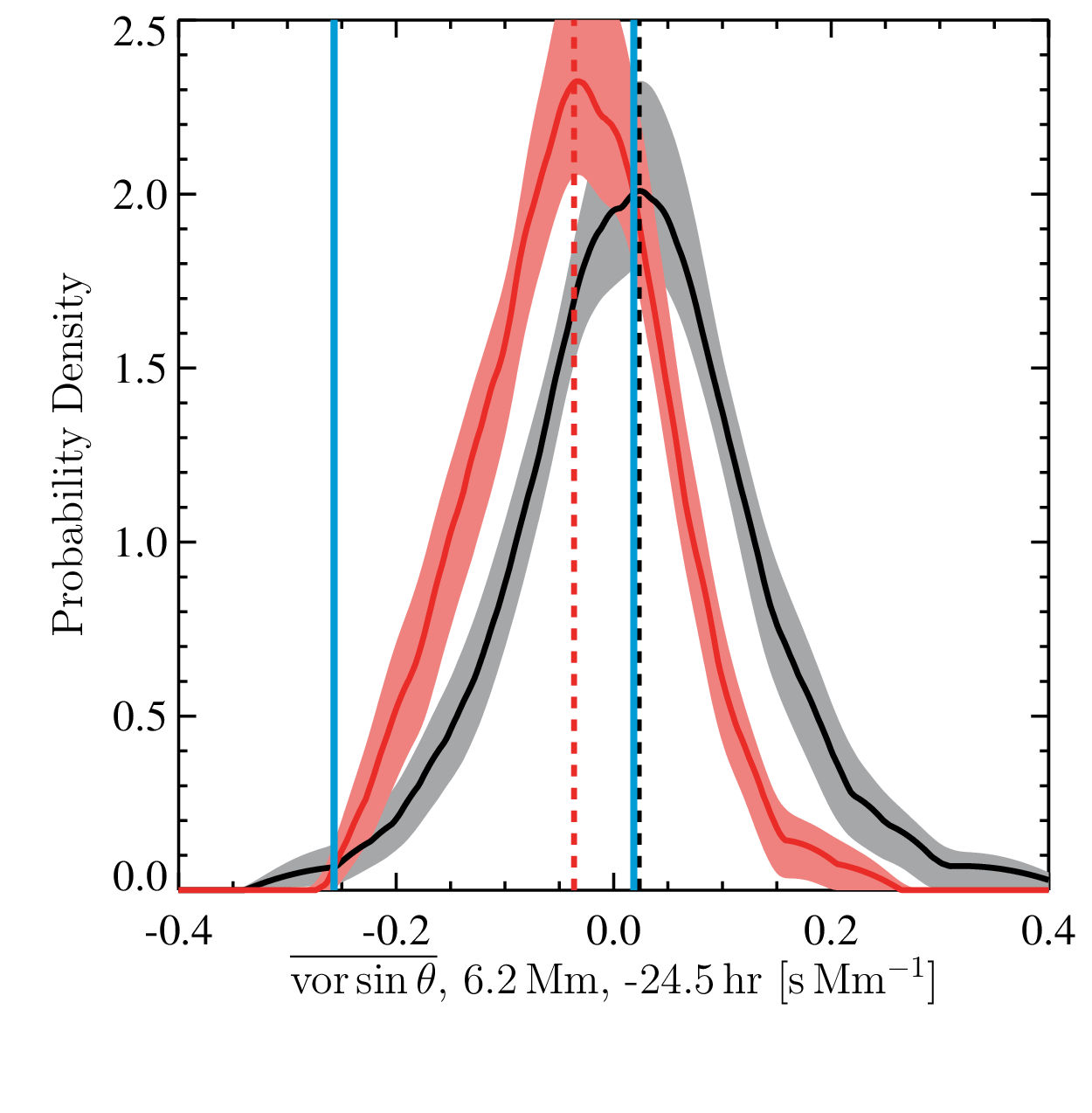}
{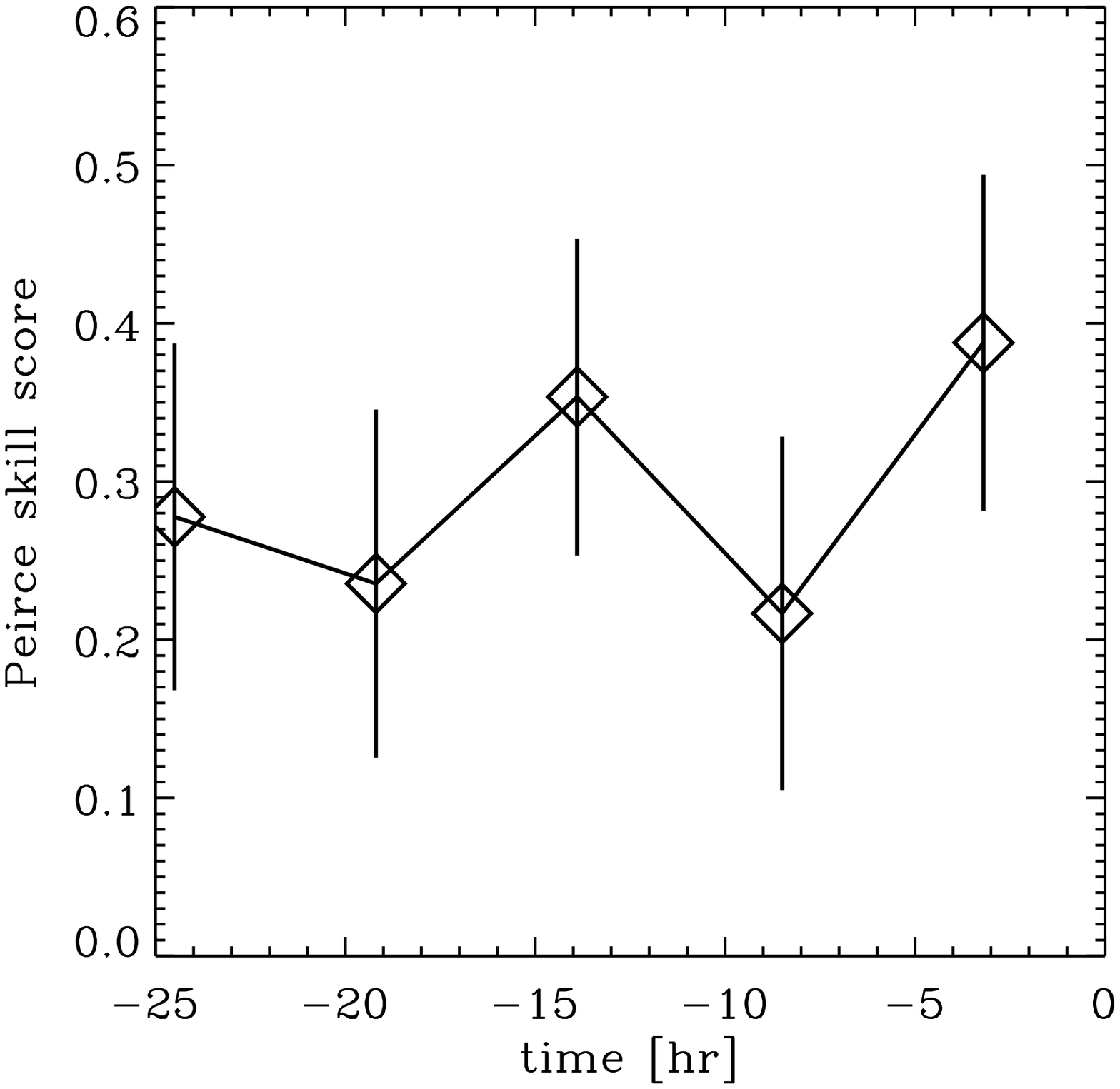}
\caption{Nonparametric discriminant analysis for antisymmetric averages of the
north-south travel-time differences, $\DTYS$, in filter TD3 (top), and for the
vertical vorticity, $\VORS$, in filter TD4 (bottom), for a subset of regions
with matched distributions of average magnetic flux, in the same format as
Fig.~\ref{fig:npda_field}.  The results are qualitatively very similar to those
seen in Fig.~\ref{fig:npda_dtxy} (bottom) and \ref{fig:npda_vors} for the full
samples.  The largest change is an increase in the uncertainties caused by the
smaller sample sizes.
} 
\label{fig:npda_dty_vors_subset}
\end{figure*}

Figure~\ref{fig:npda_dty_vors_subset} shows the probability density estimates
and the time variation of the skill score for $\DTYS$ in filter TD3 and for
$\VORS$ in filter TD4.  Qualitatively, the results are extremely similar to
those shown in Figures~\ref{fig:npda_dtxy}, bottom and \ref{fig:npda_vors},
where the full sets of regions were included.  The distributions are peaked at
similar values, with similar widths, leading to discriminant boundaries in
approximately the same locations.  The main difference is that the uncertainty
in the skill score has increased slightly.  Thus, the ability of these
variables to distinguish PE from NE regions is not a result of a difference in
the average unsigned vertical field between the two samples, although it could
still be a result of a different aspect of the surface field (e.g., the
horizontal field, or small areas of strong vertical field).

\section{Discussion}\label{sec:discussion}

There are statistically significant differences between the properties of the
pre-emergence and non-emergence samples that persist, with relatively little
change, for at least a day prior to the onset of emergence.  However, these
differences are small, of order 1\,s or less in the travel-time shifts on
average, thus none of the variables considered can clearly distinguish an
emergence from a non-emergence for any single region (c.f.~Figs.~2 and 3 from
\citetalias{Birchetal2013}).  This is quite different from the results of
\cite{Ilonidisetal2011}, who found much larger travel-time reductions,
although that study considered waves that propagate much deeper than were
considered here.

The average unsigned magnetic flux at the surface was the best discriminator
between the two samples.  This could be a result of the appearance of small
amounts of flux at the surface, starting at least one day prior to our
definition of emergence time.  The MDI instrument is unable to resolve this
flux in a single magnetogram, and thus it is only distinguishable when
averaging over many regions.  Simulations of flux emergence
\citep[e.g.,][]{Cheungetal2010,Steinetal2011} do exhibit this type of behavior,
thus our investigation shows some support for these simulations.

It is also possible that the ability of the average magnetic flux to
distinguish the two samples is a result of a bias in the samples, either of
solar origin, or as a result of our selection criteria.  Our selection of NE
regions (see \citetalias{Lekaetal2013}) imposed a maximum field strength
allowed that was not similarly applied to the PE regions.  This could have
resulted in a bias between the two samples in the average flux.  However, it is
also possible that the difference in average flux is a result of the tendency
for active regions to emerge in the same location as prior active regions
\citep[e.g.,][]{PojogaCudnik2002}, and not directly related to the emergence
process.

While such considerations are important if the goal is to use helioseismology
to predict the emergence of active regions, our goal is simply to determine if
there is a helioseismic signal of emergence.  The helioseismic measures that best
distinguish the pre-emergence from the non-emergence regions are mean
travel-time shifts, particularly immediately prior to emergence, antisymmetric
averages of north-south and east-west travel-time differences, and an
antisymmetric average of the vertical vorticity.  The mean travel-time shifts
are correlated with the presence of surface field, and thus may not be related
to subsurface properties of the emergence.  This was confirmed by the reduced
ability of mean travel-time shifts to distinguish PE from NE regions for
subsets of the initial samples of PE and NE regions that had matched
distributions of average unsigned magnetic flux.

The signals in the north-south and east-west travel-time differences, and the
signal in the vertical vorticity appear to not be sensitive to the surface
field.  Thus, we believe there are differences in the subsurface flows that can
be detected by helioseismology prior to the emergence of significant magnetic
flux.  A converging flow could qualitatively explain the signals seen in the
north-south and east-west travel-time differences, but it appears that the flow
pattern is not a simple converging flow.  A prograde flow below the site of the
emergence would produce the observed pattern in the vertical vorticity.  This
is perhaps related to the ``small shearing flow feature'' at a depth of 2\,Mm
described by \cite{KosovichevDuvall2008} for AR\,10488.  There is no clear
evidence for a retrograde flow, as would be expected from typical rising flux
tube simulations \citep[e.g.,][]{Fan2008}, although these simulations end
approximately 20\,Mm below the surface.  Instead, we found the vertical
vorticity to be consistent with a prograde flow, and the difference in
north-south travel-time differences is comparable to that in east-west
travel-time differences, as would result from a converging flow.  Thus our
results suggest that the properties of simulated rising flux tubes must change
as they approach the surface, at shallower depths than are presently simulated,
if this is the mechanism by which active regions form.

As noted in \citetalias{Lekaetal2013}, there are several ways in which a future
investigation could improve on the present method.  However, we have already
found subtle but significant differences in the helioseismic signals from our
samples of pre-emergence and non-emergence regions that suggest a detectable
subsurface manifestation of active region formation prior to the appearance of
significant surface magnetic flux.  Our statistical results place strong
constraints on models of active region formation.

\acknowledgements

The authors acknowledge support from NASA contracts NNH07CD25C and NNH12CF23C.
This work utilizes data obtained by the Global Oscillation Network Group (GONG)
Program.  GONG is managed by the National Solar Observatory, which is operated
by AURA, Inc.\ under a cooperative agreement with the National Science
Foundation.  The data were acquired by instruments operated  by the Big Bear
Solar Observatory, High Altitude Observatory, Learmonth Solar Observatory,
Udaipur Solar Observatory, Instituto de Astrof{\'i}sica de Canarias, and Cerro
Tololo Interamerican Observatory.  MDI data were provided by the SOHO/MDI
consortium; SOHO is a project of international cooperation between ESA and
NASA.  We would like to thank I.~Gonz{\'a}lez Hern{\'a}ndez, B.~Javornik, and
T.~Dunn for their work in preparing the data base. The Discriminant Analysis
code was originally developed with funding from AFOSR under contracts
F49620-00-C-0004 and F49620-03-C-0019.  ACB acknowledges DFG SFB 963
``Astrophysical Flow Instabilities and Turbulence'' (project A18) aimed at
understanding solar and stellar dynamos. 

\appendix

\section{The Influence of Duty Cycle}\label{sec:duty}

\begin{figure*}
\plottwo{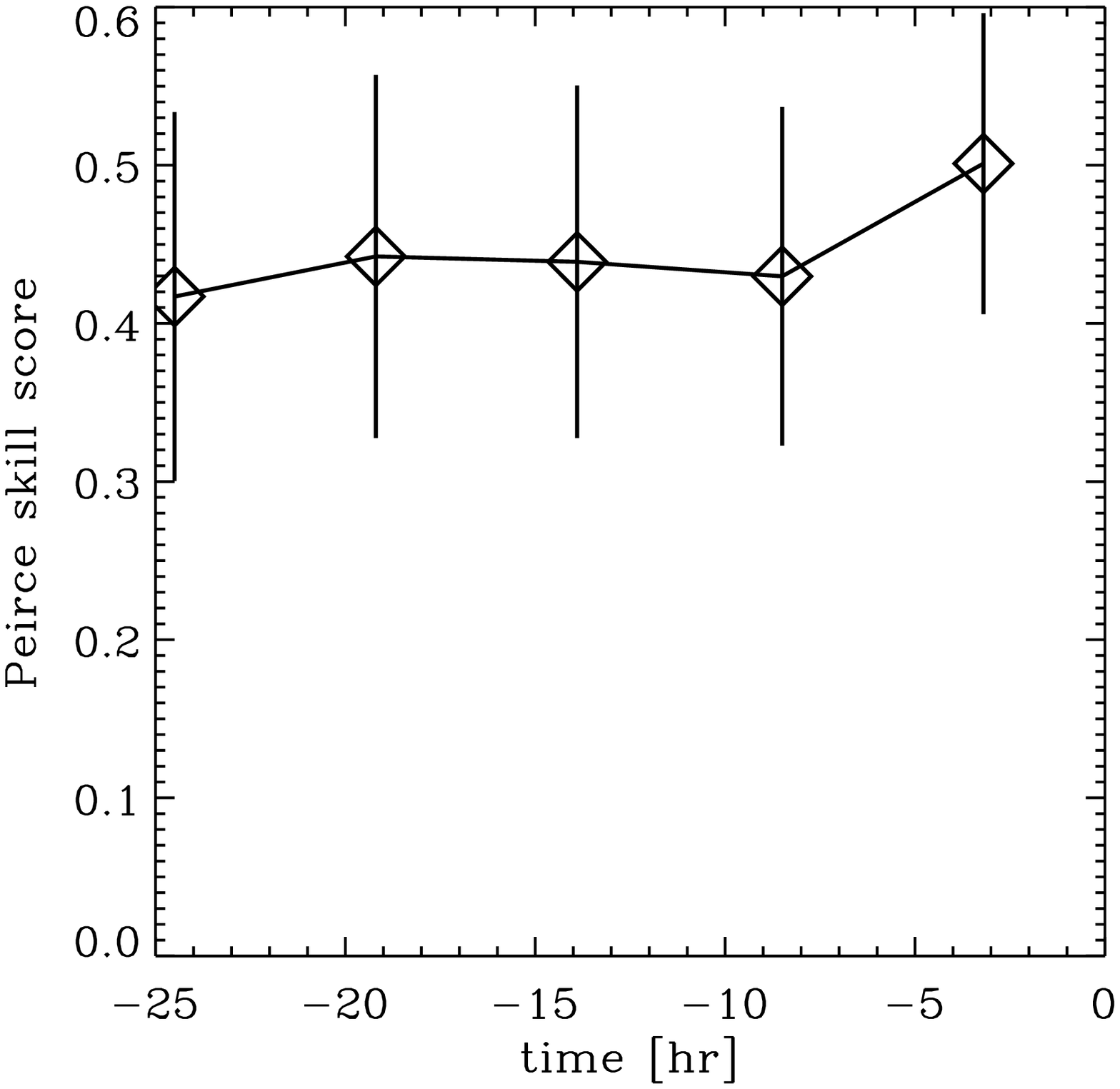}
{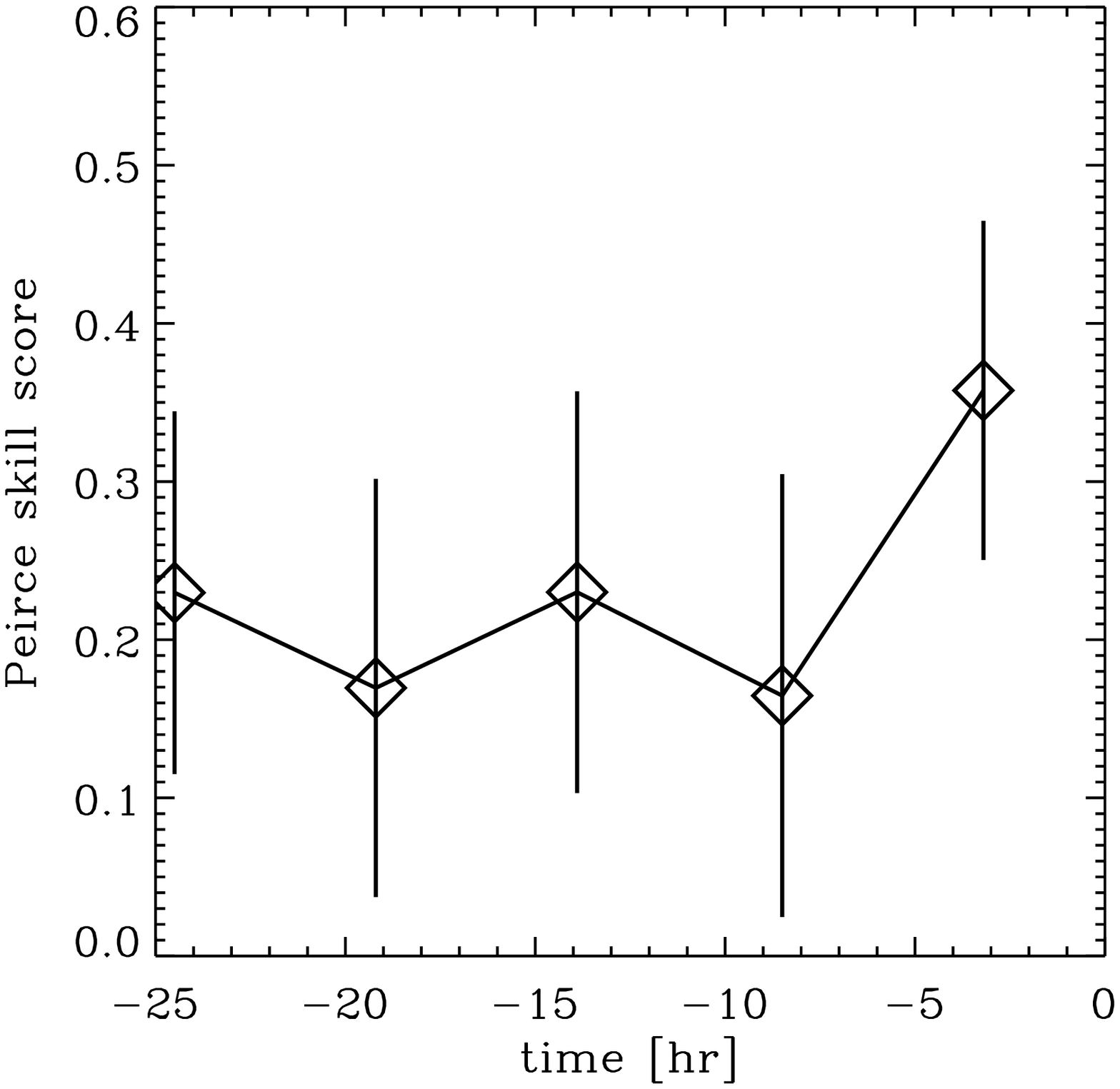}

\plottwo{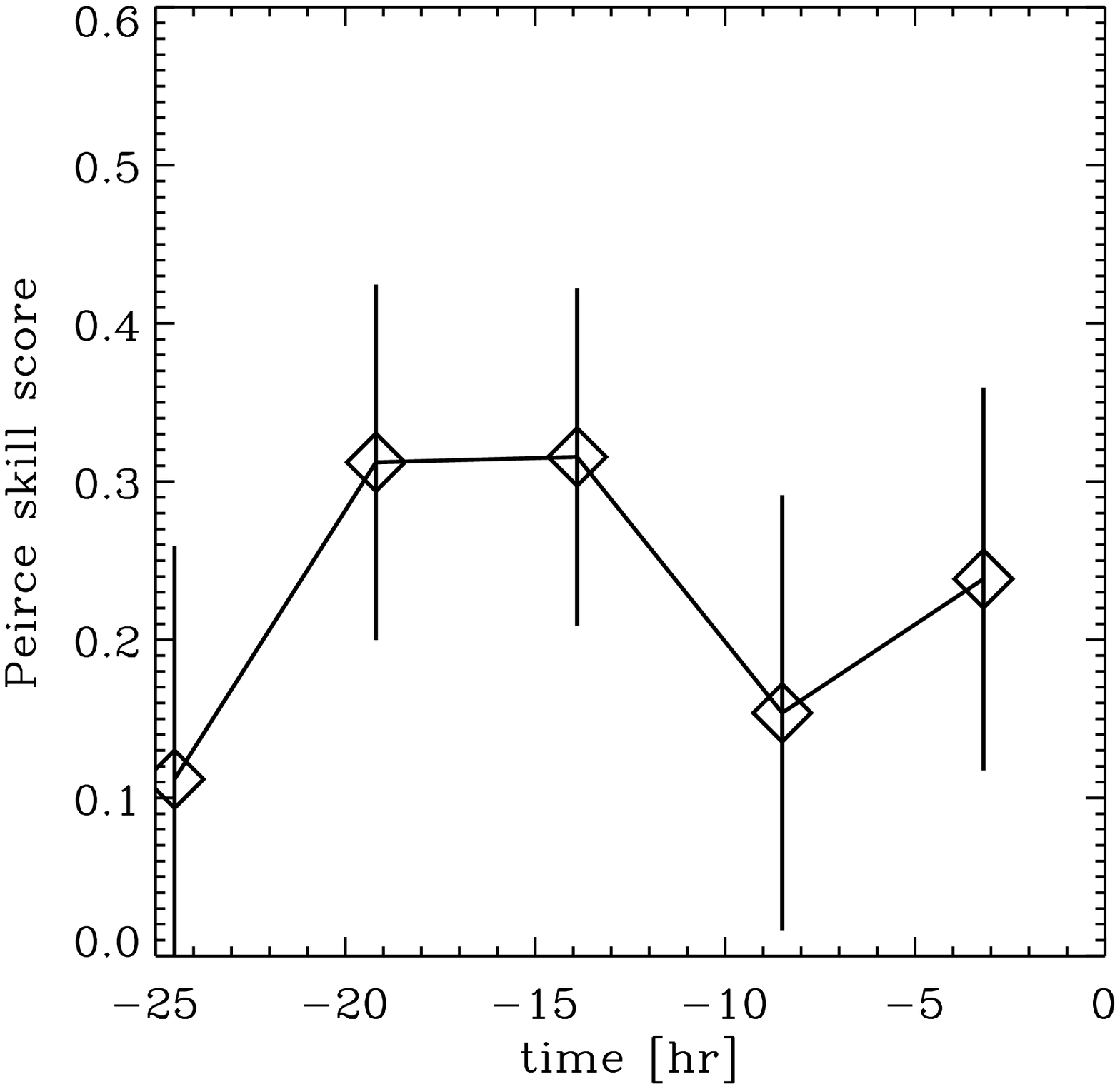}
{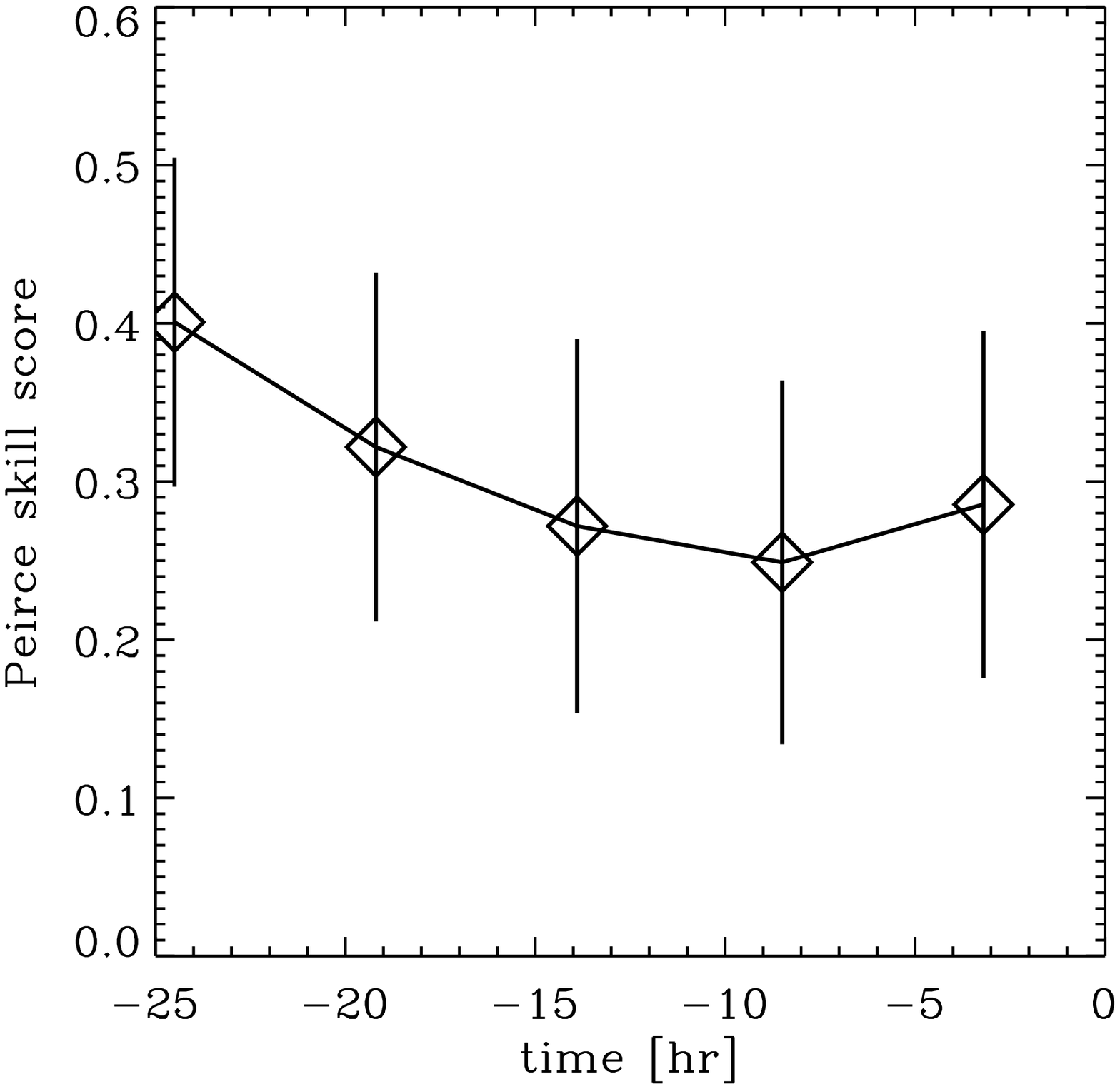} 
\caption{Evolution of the Peirce skill score for four variables: {\it top,
left}: $\UFIELD$, {\it top, right}: $\DTMEAN$ in filter TD5, {\it bottom,
left}: $\DTXC$ in filter TD3, and {\it bottom, right}: $\DTYS$ in filter TD3.
Only regions with good data for all time intervals and seismology variables
were used.  Compared to the right panels of Figs.~\ref{fig:npda_field},
\ref{fig:npda_dtmean}, \ref{fig:npda_dtxy}, in which all available regions were
used in each plot, the same trends (or lack of trends) are seen, but with
larger uncertainties.
} 
\label{fig:segs_allgood}
\end{figure*}

Because of the varying duty cycle, some regions are only present in a subset of
the time intervals.  This could potentially influence the results, if there is
a handful of regions (with varying duty cycle) that are easy to classify.  To
check this, the analysis was repeated for the subset of regions that had a
good duty cycle for {\em every} time interval.  This severely reduces the
sample sizes, to 48 and 45 for PE and NE respectively.  The main impact of this
is an increase in the error bars, which is expected from the reduction in the
sample sizes, without greatly changing the results.  To illustrate this, the
right panels of Figures~\ref{fig:npda_field}, \ref{fig:npda_dtmean}, and
\ref{fig:npda_dtxy} have been reproduced in Figure~\ref{fig:segs_allgood} with
this subset.  All the other plots exhibit the same behavior, thus we believe
that this does not affect any of our conclusions.

\section{Monte-Carlo Experiment}\label{sec:MC}

\begin{figure*}
\plottwo{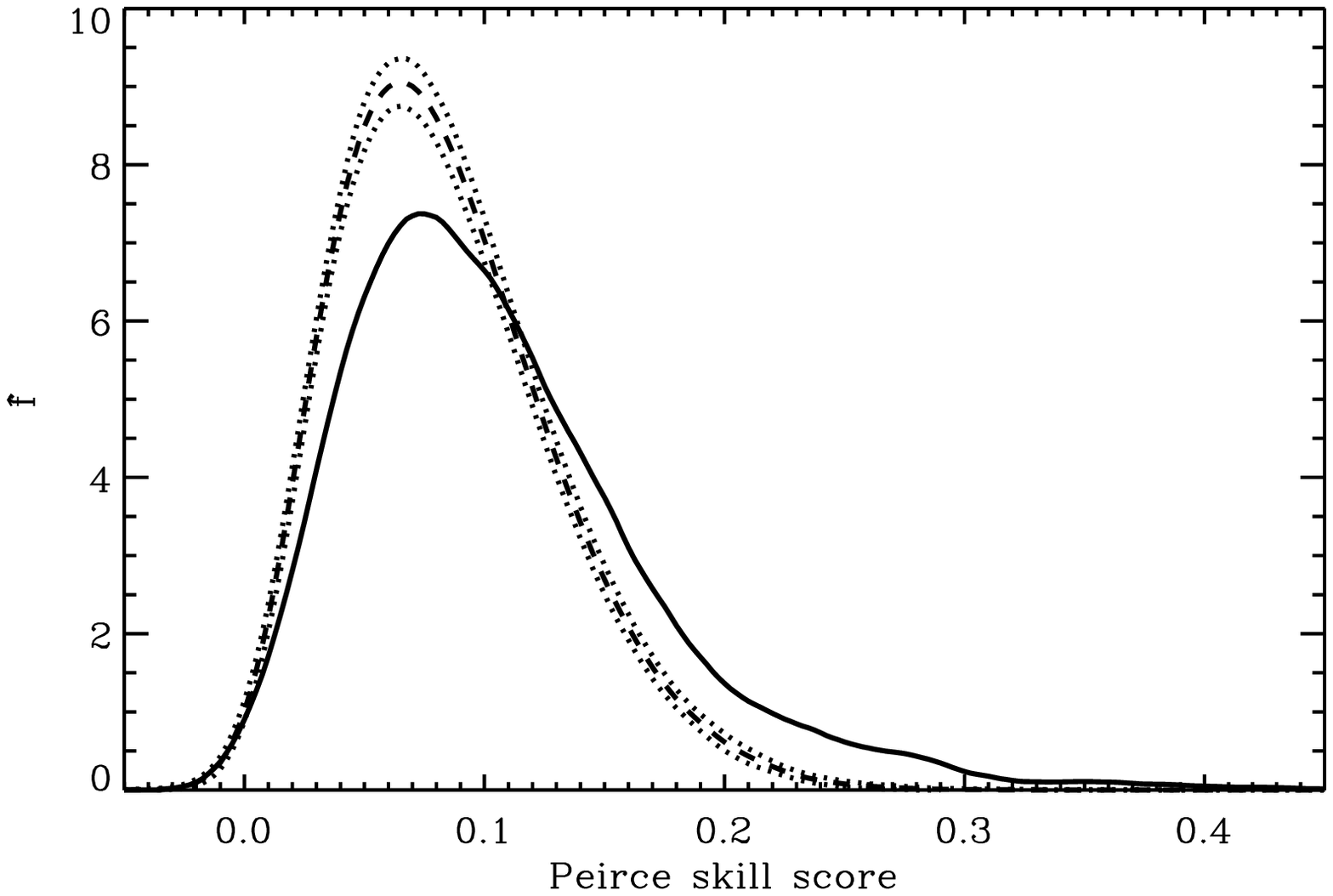}
{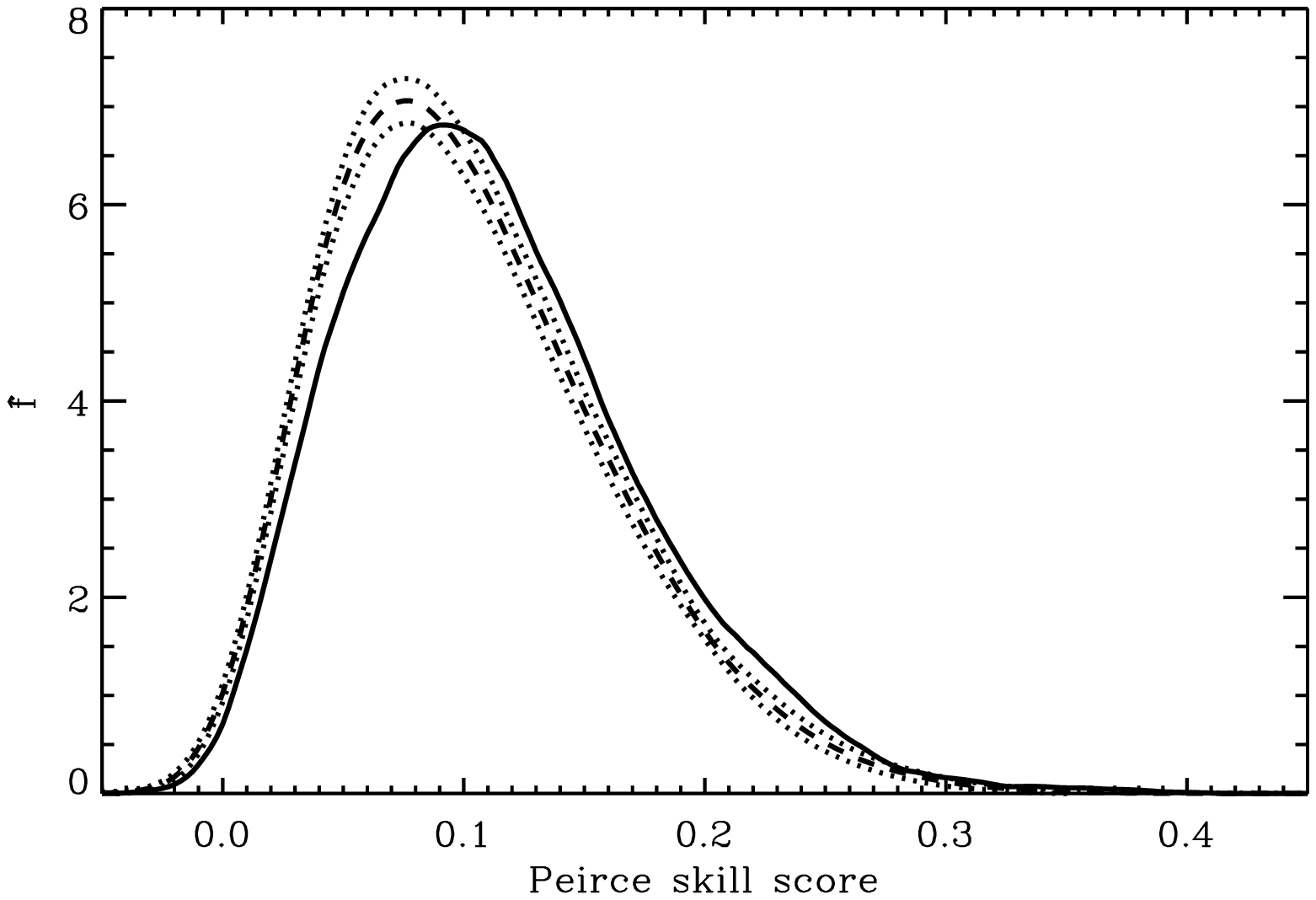}
\caption{Probability density of Peirce skill scores from the Monte Carlo
experiment (dashed curve) with $1-\sigma$ error estimate (dotted curves) and
from the variables considered for active region emergence (solid curve), using
a bootstrap estimate.  The left panel shows the results for sample sizes of 85;
the right panel shows the results for sample sizes of 50.  There is a clear
tail of the distribution of the active region emergence variables to larger
skill scores not present for the random variables, indicating that it is very
unlikely that chance alone accounts for the performance of the best variables
at distinguishing PE from NE regions.  For the sample size of 85, the variables
with skill scores above 0.27 almost certainly have a real ability to
discriminate between PE and NE regions; for the sample size of 50, it is
difficult to determine if any specific helioseismology variable has a real
ability to discriminate between PE and NE regions, but the number of
helioseismology variables with large skill scores ($\ga 0.2$) suggests that
some can discriminate between the two.
}
\label{fig:MCboot}
\end{figure*}

Given the number of variables considered compared to the number of data points,
one should ask the question: are these results simply a statistical fluke?  To
answer this, a Monte Carlo experiment was performed.  To represent one
variable, two random samples of 85 points (typical of the sample sizes in
\S\ref{sec:results} and Table~\ref{tbl:peirce_boot_npda_equal_n1_seg}) each
were drawn from the same normal distribution.  This was repeated for 66,250
variables (50 times the number of active region emergence variables), changing
only the random number seed between variables.  Nonparametric discriminant
analysis was applied to the resulting values, and an unbiased bootstrap
estimate of the Peirce skill score was made for each variable.  The
distribution of the resulting skill scores is shown in Figure~\ref{fig:MCboot},
left, along with the distribution of the variables considered for active region
emergence.  Compared with the random variables, there is a preponderance of
large skill score values for the emergence variables.  Dividing the random
variables into 50 sets of size equal to the number of emergence variables shows
that the typical maximum skill score achieved is about 0.27, so the probability
of getting a skill score greater than that if there is no information in the
variable is extremely small.  However, the distribution of random variables and
of active region emergence variables show considerable overlap below a skill
score of about 0.2.

To compare with the results when the distribution of magnetic flux was matched
between NE and PE regions (\S\ref{sec:fluxmatch} and
Table~\ref{tbl:subsetpeirce_boot_npda_equal_n1_seg}), the experiment was
repeated for two random samples of 50 points each.  The resulting distribution
of skill scores is shown in Figure~\ref{fig:MCboot}, right.  There is still a
preponderance of large skill scores for the emergence variables compared to the
random variables, but it is less pronounced than for the larger sample size.
Again dividing the random variables into 50 sets of size equal to the number of
emergence variables shows that the typical maximum skill score achieved is now
about 0.34, but was as high as 0.43.  In this case, it is no longer possible to
determine whether any individual variable has any real ability to discriminate
between PE and NE regions.  However, it is possible to infer that there are
more variables with high skill scores than would be expected from chance alone.
The number of variables with skill score $\ge 0.27$ (the value used in
Table~\ref{tbl:peirce_boot_npda_equal_n1_seg}) is 17, compared with an expected
number of 11. The chance of getting at least this many variables by chance is
approximately 2\%.  The number of variables with skill score $\ge 0.2$ is 120,
compared with an expected number of 86. The chance of getting at least this
many variables by chance is less than 2\%.  Thus there is reason to believe
that there is a difference in the helioseismology variables between PE and NE
regions.

For the results presented here, the distributions were assumed to be normal, as
this is a reasonable approximation to the expected noise distribution.
However, to confirm the results, the Monte Carlo experiments were also
performed drawing at random from a Cauchy distribution and from a cosine
distribution.  For the Cauchy distribution, which has longer tails than a
normal distribution, the largest skill scores obtained were less than the
largest skill scores for the normal distribution.  For the cosine distribution,
which has shorter tails than a normal distribution, the largest skill scores
were very similar to those obtained from the normal distribution.  In all
cases, there is a clear tail of the distribution of the emergence variables to
larger skill scores not present for the random variables, indicating that it is
very unlikely that chance alone accounts for the performance of the best
variables at distinguishing PE from NE regions.

\end{document}